\documentclass[a4paper, amsfonts, amssymb, amsmath, reprint, showkeys, nofootinbib, twoside, preprintnumbers]{revtex4-1}
\usepackage[english]{babel}
\usepackage[utf8]{inputenc}
\usepackage[colorinlistoftodos, color=green!40, prependcaption]{todonotes}
\usepackage{amsthm}
\allowdisplaybreaks
\usepackage{mathtools}
\usepackage{physics}
\usepackage{xcolor}
\usepackage{graphicx}
\usepackage[left=23mm,right=13mm,top=35mm,columnsep=15pt]{geometry} 
\usepackage{adjustbox}
\usepackage{placeins}
\usepackage[T1]{fontenc}
\usepackage{lipsum}
\usepackage{csquotes}
\usepackage{verbatim}
\usepackage[pdftex, pdftitle={Article}, pdfauthor={Author}]{hyperref} 
\usepackage{aas_macros}
\usepackage{multirow}
\begin{document}
\preprint{PI/UAN-2023-723FT}
\title{Particle-like solutions in the generalized SU(2) Proca theory}
\author{Jhan N. Martínez}
    \email[Correspondence email address: ]{jhamarlo@correo.uis.edu.co}
    \affiliation{Escuela de F\'{\i}sica, Universidad Industrial de Santander,  Ciudad Universitaria, Bucaramanga 680002, Colombia\\}
\author{José F. Rodr\'iguez }
    \email[Correspondence email address: ]{jose.rodriguez2@correo.uis.edu.co}
    \affiliation{Escuela de F\'{\i}sica, Universidad Industrial de Santander,  Ciudad Universitaria, Bucaramanga 680002, Colombia\\}
    \affiliation{ICRANet, Piazza della Repubblica 10, 65122, Pescara PE, Italy\\}
\author{Yeinzon Rodr\'iguez }
    \email[Correspondence email address: ]{yeinzon.rodriguez@uan.edu.co}
    \affiliation{Escuela de F\'{\i}sica, Universidad Industrial de Santander,  Ciudad Universitaria, Bucaramanga 680002, Colombia\\}
    \affiliation{ Centro de Investigaciones en Ciencias B\'asicas y Aplicadas, Universidad Antonio Nari\~no,Cra 3 Este \# 47A - 15, Bogot\'a D.C. 110231, Colombia\\}
\author{Gabriel G\'omez}
    \email[Correspondence email address: ]{gabriel.gomez.d@usach.cl}
    \affiliation{Departamento de F\'isica, Universidad de Santiago de Chile,\\Avenida V\'ictor Jara 3493, Estaci\'on Central, 9170124, Santiago, Chile}


\begin{abstract}
The generalized SU(2) Proca theory is a vector-tensor modified gravity theory \textcolor{black}{where the action is  invariant under both diffeomorphisms and global internal transformations of the SU(2) group.} This work constitutes the first approach \textcolor{black}{to investigate the physical properties of the theory at astrophysical scales.} 
We have found solutions that \textcolor{black}{naturally} generalize the particle-like solutions of the Einstein-Yang-Mills equations, also known as gauge boson stars. 
\textcolor{black}{Under the requirement that the solutions must be static, asymptotically flat, and globally regular, the t'Hooft-Polyakov magnetic monopole configuration for the vector field rises as one viable possibility}. The solutions \textcolor{black}{have been obtained} analytically through asymptotic expansions and numerically by solving the boundary value problem. We have found new features in the solutions such as regions with negative effective energy density and imaginary effective charge. We have also obtained a new kind of globally charged solutions for some \textcolor{black}{region in the parameter space} \textcolor{black}{of the theory}. Furthermore, we have constructed equilibrium sequences and found turning points in some cases. \textcolor{black}{These results hint towards} the existence of stable solutions which are absent in the Einstein-Yang-Mills case.
\end{abstract}


\maketitle

\section{Introduction} \label{sec:outline}

Although the theory of general relativity (GR) is \textcolor{black}{currently} the best description of the gravitational interaction in the local universe, see e.g. Refs. \cite{LIGOScientific:2017vwq,multime,multimes,Akiyama:2019cqa,Abuter:2018drb}, it \textcolor{black}{operates effectively at some range of energy scales, losing predictability beyond it} 
\cite{1994PhRvD..50.3874D,Burgess:2003jk}. 
\textcolor{black}{This is observed, for example, in the unavoidable prediction of singularities \cite{PhysRevLett.14.57,hawandpen} and in the perturbative non-renormalizability of the theory \cite{Deser:1974hg}.}
On the other hand, at cosmological scales, GR needs to introduce two \textcolor{black}{exotic} fluids, known as dark energy and dark matter, to \textcolor{black}{successfully} explain the dynamics of the universe \cite{1999Sci...284.1481B, 2015Sci...347.1100S}. \textcolor{black}{They are exotic because, despite the many efforts scientists have put in the latest decades to uncover their nature, this  remains a mystery \cite{amendola2010dark}.}  

\textcolor{black}{These two facts drive part of the scientific community to consider classical modified gravity models, among other plausible alternatives, as strong candidates to replace GR,} see, e.g., Ref. \cite{Shankaranarayanan:2022wbx,Heisenberg:2018vsk,Clifton:2011jh,Capozziello:2011et,Tsujikawa:2010zza}. 

\textcolor{black}{Our choice in this paper is to consider the generalized SU(2) Proca theory (GSU2P) which} is a vector-tensor modified gravity theory that introduces extra gravitational degrees of freedom. \textcolor{black}{The latter are vector fields that belong to the Lie algebra of the SU(2) group and the action is invariant under global transformations of this type.} This theory aims at generalizing the well known Proca theory, the latter being not thought as the theory of a massive spin-1 field but as a theory that breaks the internal gauge invariance.  The theory is constructed so that \textcolor{black}{the correct number of physical degrees of freedom are propagated, i.e. the Ostrogradski \textcolor{black}{instability} is absent, at least in flat spacetime \cite{GallegoCadavid:2020dho,GallegoCadavid:2022uzn,Allys:2016kbq} (see also Ref. \cite{GallegoCadavid:2021ljh} for an extended version). The GSU2P is the simplest non-Abelian version of the generalized Proca theory \cite{Tasinato:2014eka,Heisenberg:2014rta,Allys:2015sht,Allys:2016jaq,BeltranJimenez:2016rff,GallegoCadavid:2019zke} which is, in turn, the vector version of the well known scalar Galileon or Horndeski's theory \cite{1974IJTP...10..363H,deffayet1,Deffayet:2009mn,contrater} (see Ref. \cite{Rodriguez:2017ckc} for a review).}

Cosmological observations provide excellent compelling data to test modified gravity theories \cite{Planck:2018vyg,LIGOScientific:2017vwq, multime, multimes,Shankaranarayanan:2022wbx}. However, the introduction of vector fields poses a challenge in describing the large-scale structure of the universe which, as a first approximation (\textcolor{black}{according to observations}), is homogeneous and isotropic. Thus, \textcolor{black}{it is necessary, for instance, to take a large number of random vector fields so that,
on average, the resulting configuration is consistent with the background symmetry} \cite{Golovnev:2008cf} or to consider only the temporal component of just one vector field \cite{DeFelice:2016yws}. \textcolor{black}{Another successful possibility is} the purely spatial configuration known as the cosmic triad \cite{Armendariz-Picon:2004say,Golovnev:2008cf,Maleknejad:2011jw,Maleknejad:2011sq,Maleknejad:2011jw} that arises naturally \textcolor{black}{(see e.g. Ref. \cite{Maleknejad:2011jr})} in the case of an SU(2) symmetry\footnote{\textcolor{black}{A less natural realization is to consider the cosmic triad in a (broken) U(1) $\otimes$ U(1) $\otimes$ U(1) invariant theory  \cite{Golovnev:2008cf,Emami:2016ldl}.}}.  \textcolor{black}{The cosmic triad is a set of three vector fields mutually orthogonal and of the same norm that spontaneously breaks both the internal SU(2) symmetry and the external SO(3) symmetry leaving an unbroken SO(3) diagonal subgroup.} Consequently, models with such a symmetry are good candidates for describing the universe at large scales. In the case of the GSU2P theory, it has already been proved that the theory predicts periods of primordial \cite{Garnica:2021fuu} and late-time accelerated expansion \cite{Rodriguez:2017wkg}. The next immediate step is to test the theory at astrophysical scales \textcolor{black}{in the context of compact objects. This has been a compulsory programme in modified gravity theories that looks for distinctive observational signatures of these theories}.

\textcolor{black}{Neither the Einstein field equations in vacuum nor the Yang-Mills equations exhibit globally regular solutions with localized energy} (see Ref. \cite{1982PhRvD..25.2515W} and references therein). A \emph{necessary} condition for the existence of equilibrium configurations is the presence of repulsive and attractive interactions. 
Thus, \textcolor{black}{it is reasonable to study the  Einstein-Yang-Mills (EYM) scenario} to search for the existence of stationary, localized, and non-singular solutions. We will refer to such solutions as {\em particle-like or soliton solutions} (see, e.g., Ref. \cite{1999PhR...319....1V} for a review).  \textcolor{black}{We consider the solutions we have found as classical solitons since their energy density is localized and have finite positive total mass-energy, i.e., they resemble extended particles.}
Beyond purely academic motivations, these objects can be of great interest from an astrophysical point of view as they can serve as models of dark matter\footnote{They can contribute partially to the dark matter content of the universe (for dark matter candidates see, e.g., Refs. \cite{Roszkowski:2017nbc,Liebling:2012fv} and references therein).}  or black holes mimickers \cite{2017LRR....20....5L}. \textcolor{black}{Among early works, it is of particular interest the work of R. Bartnik and J. McKinnon \cite{1988PhRvL..61..141B}, who found a family of particle-like solutions of the EYM equations. However, it was shown later that these solutions are unstable against radial perturbations \cite{1990PhLB..237..353S}.} 

Immediate generalizations of the EYM particle-like solutions were the Einstein-SU(2) $\otimes$ SU(2) (Skyrmion) 
\cite{1991PhLB..271...61H}, the Einstein-SU(2) Proca  and the Einstein-Yang-Mills-Higgs  \textcolor{black}{solutions} (see Ref. \cite{1993PhRvD..47.2242G} for the latter two proposals). Indeed, it was found that one branch of the equilibrium sequence of the Skyrmion solution is stable. On the other hand, black hole solutions for all the previous cases were also found\footnote{Surprisingly, the Skyrme black hole was found before the EYM particle-like solution.} \cite{1990PhRvL..64.2844B,1986PhLB..176..341L, 1993PhRvD..47.2242G}. Moreover, it was found that one branch of the Skyrme black hole and one branch of the \textcolor{black}{Einstein-}SU(2) Proca black hole are stable and share similar properties. \textcolor{black}{The stability was related to the fact that the rest mass-energy term dominates over the kinetic energy \cite{1995PhRvD..51.1510T}. Therefore, additional energy contributions to the effective rest mass term can eventually stabilize the solutions.} Finally, asymptotic de-Sitter and anti-de Sitter generalizations can be found in Refs. \cite{2000PhRvD..62d3513B, 2016PhRvD..94d4048P}.

It is essential to mention that \textcolor{black}{employing} SU(2) vector fields are not the only \textcolor{black}{way} to find self-gravitating solitons. For example, the first boson stars in GR involve a complex scalar field \cite{1968PhRv..172.1331K, 1969PhRv..187.1767R}. Moreover, a massive Abelian spin-1 boson minimally coupled to Einstein's gravity can form solitons \cite{2016PhLB..752..291B}. The non-minimal \textcolor{black}{coupling to gravity} versions of the previous cases have also been discovered \cite{2016PhRvD..93l4057B, 2017PhRvD..95b4027B, 2017PhRvD..96d4017M}. One non-minimal \textcolor{black}{coupling to gravity} version of the SU(2) case has been studied in Refs. \cite{2007PhLB..644..294B, 2016PhRvD..93b4008B,2016PhRvD..93h4004B}, where the coupling is introduced through a term involving the double dual Riemann tensor. This term induces unavoidably both Laplacian and ghost instabilities \textcolor{black}{on its own. Although such instabilities were found in a Friedmann–Lemaitre–Robertson–Walker background \cite{BeltranJimenez:2013btb}, it is very likely that  things do not change significantly in other background metrics unless other terms are added as a part of the gravity theory to ensure its viability.} In addition, Refs. \cite{2007PhLB..644..294B, 2016PhRvD..93b4008B,2016PhRvD..93h4004B} used  the \emph{Wu-Yang} monopole configuration which is a special case of the one we are going to employ here\footnote{\textcolor{black}{The Wu-Yang monopole corresponds, indeed, to the most trivial configuration among those studied in this paper.}}.

In this work, we have taken the first step to test the GSUP2 theory at astrophysical scales. We have focused on the static, spherically symmetric, and asymptotically flat case, which is of significant astrophysical interest. We have chosen the \emph{t'Hooft-Polyakov magnetic monopole} ansatz which is the same configuration used in the original EYM soliton. We have studied analytically the behaviour of the equations near the origin and at infinity. In addition, we have performed a numerical analysis of the solutions. \textcolor{black}{We have also constructed equilibrium sequences and found turning points, the latter being possible indicators of changes in the stability properties of the solutions.} We have paid attention to the energy density arising from the new terms in the action whose contribution can be negative. \textcolor{black}{We have found that, with the exception of one of its Lagrangian pieces, the GSU2P gives rise to generalized EYM particle-like solutions.} 

This work is organized as follows. Section II describes the model we have used and presents the tensor field equations (see Appendix \ref{sec:appendixFE}). Section III introduces the stationary and spherically symmetric configurations for the metric and the vector fields; \textcolor{black}{in addition, analytic solutions for some of the Lagrangian pieces of the theory are presented; furthermore, the boundary conditions to obtain asymptotically regular flat solutions and to analyze the  behaviour near the origin and at infinity are described. Section IV presents the numerical solutions satisfying the boundary conditions; the new features of the solutions, such as a global charge and negative energy regions, are exposed; moreover, equilibrium sequences are constructed and some brief arguments about the stability of the solutions are given. Finally, in Section V, the results and some observational perspectives of this work are discussed.} Throughout this article, we have used geometrized units ($c = G = 1$), where $c$ is the speed of light and $G$ is the universal gravitational constant. Greek indices represent space-time indices and run from 0 to 3. Latin indices represent SU(2) group indices and run from 1 to 3. We have used the sign convention $(+,+,+)$ according to Ref. \cite{2017grav.book.....M}.

\section{Generalized SU(2) Proca Theory}

The action \textcolor{black}{that defines this theory was introduced originally in Ref. \cite{Allys:2016kbq} and later reconstructed in Ref. \cite{GallegoCadavid:2020dho} (see also Ref. \cite{GallegoCadavid:2022uzn}) to take into account the whole constraint algebra that avoids the propagation of unphysical degrees of freedom (see Refs. \cite{ErrastiDiez:2019trb,ErrastiDiez:2019ttn}). In this paper, we have made use of the GSU2P in the form given in Ref. \cite{Garnica:2021fuu}:}
\begin{equation}
     S = \frac{1}{16\pi}\int R \ \sqrt{-g}\ d^4x + \frac{1}{16\pi}\int \mathcal{L} \  \sqrt{-g} \ d^4x \,,
     \label{eqn:action}
\end{equation}
where $g$ is the determinant of the metric, $R$ is the Ricci scalar, and
\begin{align}
\begin{aligned}
\mathcal{L}=-& F_{\mu \nu}^{a}F^{\mu \nu}_{a}-2\mu^2 B^{\mu}_{a}B_{\mu}^{a} \\
&+\alpha_1\left(\mathcal{L}^1_{4,2}-2\mathcal{L}^4_{4,2}-\frac{20}{3}\mathcal{L}^5_{4,2}+5\mathcal{L}^{7}_{2}\right)\\
&+ \alpha_3\Big(2\mathcal{L}^2_{4,2}+\mathcal{L}^3_{4,2}+\frac{7}{20}\mathcal{L}^4_{4,2}+\frac{14}{3}\mathcal{L}^5_{4,2}\\
&-8\mathcal{L}^6_{4,2}+\mathcal{L}^{7}_{2} \Big)\\
&+ \chi_1 \mathcal{L}^{1}_{2}+\chi_2\mathcal{L}^{2}_{2} +\chi_4\left(\mathcal{L}^{4}_{2}-\frac{\mathcal{L}^{7}_{2}}{2}\right)+\chi_5\mathcal{L}^{5}_{2}\\
&+ \chi_6\left(\mathcal{L}^{6}_{2}-3\mathcal{L}^{7}_{2}\right),
\label{eqn:lagrangian}
\end{aligned}
\end{align}
where $\mu \equiv m_a/\hbar$, $m_a$ being the mass of the vector field $B_\mu^a$ \textcolor{black}{(the same mass value for all three fields has been chosen) and $\hbar$ being the reduced Planck constant}, $F_{\mu \nu}^{a}\equiv \partial_{\mu}B_{\nu}^{a}-\partial_{\nu}B_{\mu}^{a}+\tilde{g}\epsilon^{a}_{\hspace{2mm}bc}B_{\mu}^{b}B_{\nu}^{c}$ \textcolor{black}{is the gauge field strength tensor}, with \textcolor{black}{$\tilde{g}$} being the gauge charge and \textcolor{black}{$\epsilon^{a}_{\hspace{2mm}bc}$ being the structure
constant tensor of the SU(2) group}, and the $\alpha_i$ and $\chi_i$ are free parameters of the theory. The explicit form of the Lagrangian pieces is given by
\begin{align}
\begin{aligned}
\mathcal{L}^{1}_{4,2}\equiv&\left(B_{b}\cdot B^{b}\right)\left[S^{\mu a}_{\mu}S_{\nu a}^{\nu}-S^{\mu a}_{\nu}S_{\mu a}^{\nu}\right]\\
&+2 \left(B_{a}\cdot B_{b}\right)\left[S^{\mu a}_{\mu}S_{\nu }^{\nu b}-S^{\mu a}_{\nu }S_{\mu }^{\nu b}\right] \,,
\label{}
\end{aligned}
\end{align}
\begin{align}
\begin{aligned}
\mathcal{L}^{2}_{4,2}\equiv&A_{\mu \nu}^{a}S^{\mu b}_{\sigma}B^{\nu}_{a}B^{\sigma}_{b}-A_{\mu \nu}^{a}S^{\mu b}_{\sigma}B^{\nu}_{b}B^{\sigma}_{a}\\
&+A_{\mu \nu}^{a}S^{\sigma b}_{\sigma}B^{\mu}_{a}B^{\nu}_{b} \,,
\label{}
\end{aligned}
\end{align}
\begin{align}
\begin{aligned}
\mathcal{L}^{3}_{4,2}\equiv&B^{\mu a}R^{\alpha}_{\hspace{2mm}\sigma \rho \mu}B_{\alpha a}B^{\rho c}B^{\sigma}_{c}\\
&+\frac{3}{4}\left(B^{a} \cdot B_{a}\right)\left(B_{b} \cdot B^{b}\right)R \,,
\label{}
\end{aligned}
\end{align}
\begin{align}
\begin{aligned}
\mathcal{L}^{4}_{4,2}\equiv&\Big[\left(B^{a}\cdot B_{a}\right)\left(B^{b}\cdot B_{b}\right)\\
&+2\left(B^{a}\cdot B^{b}\right)\left(B_{a}\cdot B_{b}\right)\Big]R \,,
\label{}
\end{aligned}
\end{align}
\begin{equation}
\mathcal{L}^{5}_{4,2} \equiv G_{\mu \nu}B^{\mu a}B^{\nu}_{a}\left(B^{b} \cdot B_{b}\right) \,,
\label{}
\end{equation}
\begin{equation}
\mathcal{L}^{6}_{4,2}\equiv G_{\mu \nu}B^{\mu a}B^{\nu b}\left(B_{a} \cdot B_{b}\right) \,,
\label{}
\end{equation}
\begin{equation}
\mathcal{L}^{1}_{2} \equiv (B^{a} \cdot B_{a}) (B^{b} \cdot B_{b}) \,,
\label{}
\end{equation}
\begin{equation}
\mathcal{L}^{2}_{2} \equiv (B^{a} \cdot B_{b}) (B^{b} \cdot B_{a}) \,,
\label{}
\end{equation}
\begin{equation}
\mathcal{L}^{3}_{2} \equiv B_{\mu}^{b} B_{\rho b} A^{\mu \nu a} A^{\rho}_{\;\; \nu a} \,,
\label{}
\end{equation}
\begin{equation}
\mathcal{L}^{4}_{2} \equiv B_{\mu}^{b} B_{\rho a} A^{\mu \nu a} A^{\rho}_{\;\;\nu b} \,,
\label{}
\end{equation}
\begin{equation}
\mathcal{L}^{5}_{2} \equiv B_{\mu a} B_{\rho}^{b} A^{\mu \nu a} A^{\rho}_{\;\; \nu b} \,,
\label{}
\end{equation}
\begin{equation}
\mathcal{L}^{6}_{2} \equiv (B^{b} \cdot B_{b}) A_{\mu \nu a} A^{\mu \nu a} \,,
\label{}
\end{equation}
\begin{equation}
\mathcal{L}^{7}_{2} \equiv (B^{b} \cdot B_{a}) A_{\mu \nu b} A^{\mu \nu a} \,,
\label{}
\end{equation}
where $A^{a}_{\mu \nu} \equiv \nabla_{\mu}B_{\nu}^{a} - \nabla_{\nu}B_{\mu}^{a}$ and $S^{a}_{\mu \nu} \equiv \nabla_{\mu}B_{\nu}^{a} + \nabla_{\nu}B_{\mu}^{a}$ \textcolor{black}{are, respectively, the Abelian version of the non-Abelian gauge 
field strength tensor and the symmetric version of $A^{a}_{\mu \nu}$,} \textcolor{black}{$R^\alpha_{\hspace{2mm} \sigma \rho \mu}$ is the Riemann tensor, and $G_{\mu \nu}$ is the Einstein tensor.}

The form of the Lagrangian in Eq. \eqref{eqn:lagrangian} was obtained by choosing a relation among \textcolor{black}{the whole set of free parameters of the theory so that the tensor perturbations on a Friedmann-Lemaitre-Robertson-Walker background have the same form as in GR at least up to second order \cite{Garnica:2021fuu}. Therefore, the propagation speed of gravitational waves is luminal and the theory is free of ghosts and Laplacian instabilities in the tensor sector.} In this way, the theory satisfies the stringent condition imposed by the observations of the gravitational wave event GW170817 and its optical counterpart GRB 170817A which are consistent with the merger of two neutron stars \cite{LIGOScientific:2017vwq,multime,multimes}. 

\textcolor{black}{We have extrapolated this tensor sector constraint down to astrophysical scales.} \textcolor{black}{However}, this form of the \textcolor{black}{action} does not \textcolor{black}{necessarily} induce  luminal gravitational wave speed near astrophysical objects since this depends on the background configuration (see e.g. Ref. \cite{BeltranJimenez:2019xxx}). Finding out whether this constraint also implies luminal propagation \textcolor{black}{at astrophysical scales} requires first obtaining the background solution and then perturbing it. This goes beyond the scope of this work so that we have left it for a future study\footnote{On the other hand, there are no observational data of gravitational waves propagating near compact objects \cite{LIGOScientific:2021izm,Kim:2020xkm}, \textcolor{black}{so the derived constraint is not strictly applicable at the concerned scale.}}.
\subsection{Tensor field equations}
The field equations governing the behaviour of the metric and the vector fields have been obtained by varying the action in Eq. \eqref{eqn:action} with respect to $g^{\mu\nu}$ and $B^{a\mu}$, respectively:
\begin{equation}
G_{\mu \nu}=8\pi T_{\mu \nu}^{\rm eff} \,,
\label{eqn:GNAP}
\end{equation}
\begin{equation}
K_{a \mu} \equiv\frac{1}{4}\frac{\delta \mathcal{L}}{\delta B^{a\mu}} =  0 \,,
\end{equation}
where
\begin{equation}
T_{\mu \nu}^{\rm effec}\equiv -\frac{1}{8\pi \sqrt{-g}}\frac{\delta (\mathcal{L}\sqrt{-g})}{\delta g^{\mu \nu}} \,.
\label{}
\end{equation}
Appendix \ref{sec:appendixFE} shows the explicit form of the tensor field equations.  

The vector fields can be regarded as hidden degrees of freedom of the gravitational interaction or can be interpreted as a dark fluid non-minimally coupled to \textcolor{black}{gravity}. We have taken the latter  \textcolor{black}{for appropriate physical interpretation} which has allowed us to define an effective energy-momentum tensor $T^{\rm eff}_{\mu\nu}$ associated to the SU(2) vector field. We stress that both interpretations are valid, but we have considered the latter interpretation as being more convenient for this work.
\section{Regular solutions} \label{sec:asymptotic}
\subsection{Spherically symmetric configurations}
The line element of a stationary and spherically symmetric spacetime in Schwarzschild coordinates is
\begin{equation}
    ds^2 = -f(r)dt^2 + h(r)^{-1}dr^2 + r^2(d\theta^2 +\sin^2\theta d\phi^2) \,, \label{eqn:metricI}
\end{equation}
where $f(r)=e^{-2\delta(r)}h(r), \delta(r)$, and $h(r)$ are functions of the coordinate $r$. 

We require that the spacetime is regular everywhere. Therefore, the components of the metric at the origin must be finite, $f(0)<\infty$, $h(0)<\infty$, and $f'(0)=h'(0)=0$. Regularity also requires the absence of curvature divergences at $r=0$.

For convenience, the following alternative form of the metric in Eq. \eqref{eqn:metricI} has been used:  
\begin{equation}
ds^2 = -e^{2\Phi}dt^2 + \frac{1}{1-\tfrac{2m}{r}}dr^2 + r^2(d\theta^2 +\sin^2\theta d\phi^2) \,, \label{eqn:metricII}
\end{equation}
where $m \equiv r(1-h)/2$ and $\Phi\equiv\frac{1}{2}\ln (f)$.

It is of astrophysical interest to search for solutions which are asymptotically flat, thus we have imposed the  boundary conditions at spatial infinity $r\to\infty$ \textcolor{black}{so that} $\Phi\to 0$, and $m\to M={\rm const.}$.

The configuration of the vectors fields $B_{\mu}^a$ is given by the Witten \emph{ansatz} which is the more general spherically symmetric one \cite{1977PhRvL..38..121W} (see also Refs. \cite{Sivers:1986kq,Forgacs:1979zs}):
\begin{multline}
    \mathbf{B} = \frac{\tau^a}{\tilde{g}}\Biggl[A_0 \frac{x_a}{r}dt + A_1 \frac{x_a x_j}{r^2} dx^j + \frac{\phi_1}{r}\biggl(\delta_{aj} - \frac{x_a x_j}{r^2}\biggr) dx^j \\
    -\epsilon_{ajk}x^j\frac{(1+\phi_2)}{r^2}dx^k\Biggr] \,,  
\end{multline}
where $\tau_i = -i\sigma_i/2$ is a basis for \textcolor{black}{the SU(2) algebra} with $\sigma_i$ being the Pauli matrices, $A_0,\ A_1,\ \phi_1,\ \phi_2$ are functions of the coordinates $(t, r)$, $x_a$ (or $x_j$) are the space-time cartesian coordinates, and $\delta_{aj}$ is the Kronecker delta.
\textcolor{black}{Following Refs. \cite{1988PhRvL..61..141B, 1993PhRvD..47.2242G},  we have studied in this work the stationary pure magnetic configuration or t'Hooft-Polyakov monopole given by $A_0=A_1=\phi_1=0,\ \phi_2 = w(r)$}. The cartesian coordinates $x_i$ can be obtained from the polar spherical coordinates $(r, \theta,\phi)$ in the same way as in flat space. The explicit components are given by 
\begin{align} 
\tilde{g} B^1_{\mu}&= (w+1)[0 , 0 ,  \sin \phi , \sin \theta \cos \theta  \cos \phi ] \,, \nonumber\\
\tilde{g} B^2_{\mu}&=(w+1)[0 , 0 ,- \cos \phi , \sin \theta  \cos \theta  \sin \phi] \,, \nonumber\\
\tilde{g} B^3_{\mu}&=(w+1) [0 ,0 , 0 , -\sin ^2\theta ] \,.\label{eqn:tHPexplicit}
\end{align}

The gauge coupling $\tilde{g}$ has inverse units of length \textcolor{black}{in geometrized units} and we have used it to define the normalized variables $\hat{r} \equiv r\tilde{g}$ and $\hat{m}\equiv
 m \tilde{g}$. Hereafter, all the variables are dimensionless and a prime denotes differentiation with respect to $r$. For \textcolor{black}{simplicity} with the notation, we have dropped all the hats in the expressions. This implies, as we shall see later, that the theory does not fix the length or the mass scale. The results we have obtained here can be used to describe either a heavy or a light object just by choosing the gauge coupling $\tilde{g}$.
 
The effective energy-momentum tensor \textcolor{black}{allows us} to define an effective energy density measured by an observer with four velocity $u^{\mu}$,
\begin{equation}
\rho_{\rm eff}\equiv T_{\mu \nu}^{\rm eff}u^{\mu}u^{\nu} \,,
\label{eqn:rho}
\end{equation}
and an isotropic effective pressure,
\begin{equation}
p_{\rm eff}\equiv\frac{1}{3}(g^{\mu \nu}+u^{\mu}u^{\nu})T_{\mu \nu}^{\rm eff} \,.
\label{eqn:p}
\end{equation}
From these two  \textcolor{black}{definitions},  the state parameter can be defined as $\omega_{\rm eff} \equiv p_{\rm eff}/\rho_{\rm eff}$.
For simplicity, we have assumed that the energy and pressure are measured by a static observer with respect to the effective fluid, i.e., $u^{\mu} = e^{-\Phi}$. Regularity of the solution also implies that the energy and pressure are always finite.

The explicit form of the field equations in this configuration can be found in Appendix \ref{sec:appendixB}.
\subsection{Analytical solution}\label{sec:analytical}

In order to gain insight, we have solved the equations \textcolor{black}{under the assumption that the space-time norm of the gauge fields is constant},  $w=w_c={\rm const}$. We have found analytical solutions for cases where the only non-zero parameters are $\chi_1$ and $\chi_2$ \textcolor{black}{which correspond to Lagrangian pieces of the theory that involve fields only and not derivatives of them. It is natural to think hence that the specific structure of those terms helps to treat analytically the resulting vector field equation. This is not the case however for the mass term associated to the vector field as we shall see}. \textcolor{black}{The found solution corresponds interestingly to the Reissner-Nordstr\"om metric with a magnetic charge}\footnote{\textcolor{black}{Assuming a vector field configuration with just temporal components does not lead to the counterpart electric solution.}}. 

After replacing the ansatz in the vector field equation, we have obtained the following \textcolor{black}{constraint}:
\begin{equation}
    w_c(1-w_c^2)+(1+w_c)^3\chi_{12}=0 \,,
\end{equation}
where $\chi_{12} \equiv 2\chi_1 + \chi_2$. This equation has three solutions:
\begin{align}
w_c &= -1 \,,\\
w_{\rm I,II} &= \frac{1 + 2\chi_{12}\pm\sqrt{1+8\chi_{12}}}{2 -  2\chi_{12}} \label{eqn:sol_wc2} \,,
\end{align}
where the first one corresponds to the trivial case of a Schwarzschild spacetime with no vector field. 

For the solution in Eq. \eqref{eqn:sol_wc2}, the metric field equations are
\begin{equation}
    m' =  \frac{(1-w_{\rm I,II}^2)^2 -\chi_{12}(1+w_{\rm I,II})^4 }{2r^2} \,,
\end{equation}
\begin{multline}
    \Phi'= \frac{m}{r^2(1 - \frac{2m}{r})} +\frac{ \chi_{12}(1+w_{\rm I,II})^4 - (1-w_{\rm I,II}^2)^2}{2r^3(1 - \frac{2m}{r})} \,. 
\end{multline}
There are two branch solutions denoted by the subindices I and II, which correspond to the Reissner-Nordstr\"om spacetime with different effective charges:
\begin{equation}
m = M - \frac{Q_{\rm I,II}^2}{2r} \,,
\end{equation}
\begin{equation}
    \Phi = \frac{1}{2}\ln (1 - 
    \frac{2M}{r}+\frac{Q^2_{\rm I,II}}{r^2}) \,,
\end{equation}
where 
\begin{equation}
    Q^2_{\rm I,II} = \frac{1 - 4\chi_{12}(5 + 2\chi_{12}) \mp(1 + 8\chi_{12})\sqrt{1 + 8\chi_{12}}}{2(1-\chi_{12})^3} \,.\label{eqn:charge-RN-sol}
\end{equation}
and $M$ is the gravitational mass.
These solutions are degenerate with the EYM Reissner-Nordstr\"om solution \cite{1988PhRvL..61..141B}. In the $\chi_{12}\to 0$ limit, the solution I approximates in the limit to the Schwarzschild spacetime and the solution II to the  Reissner-Nordstr\"om spacetime with $Q^2 = 1$. 

\textcolor{black}{If $\chi_{12}>-1/8$, except for  $
\chi_{12}=1$ where there is a divergence, both solutions for $Q^2$ are real. For the branch I, the square of the effective charge can be negative in the interval $\chi_{12} \in (0, 1)$ i.e., gravity can be repulsive, whereas the square of the effective charge is positive for the range $(-1/8,0)\cup(1,\infty)$. For the branch II, the square of the effective charge is always positive as long as $\chi_{12} > -1/8$ and except for $\chi_{12}=1$.}

Motivated by the previous result and following Ref. \cite{1988PhRvL..61..141B}, we have defined an effective charge to characterize the metric:
\begin{equation}
Q_{\rm eff}^2 \equiv 2r(M-m) \,.
\label{eq7701}
\end{equation}
\textcolor{black}{This charge does not represent the strength of an interaction, it simply serves to compare the behaviour of the obtained metric with that of the Reissner-Nordstr\"om solution. Its value is obtained from the asymptotic behaviour of the function $m(r)$.}
In the next subsections, we will study analytically the \textcolor{black}{full} model near the origin and at spatial infinity.
\subsection{Asymptotic series solution at  the origin}

In order to find a regular solution at the origin \textcolor{black}{for the full model}, we have expressed the field variables as asymptotic series:
\begin{equation}
m = a_0 + a_1 r + a_2 r^2 + a_3r^3 + a_4r^4 + a_5r^5   +\mathcal{O}(r^6) \,,
\label{}
\end{equation}
\begin{equation}
w = b_0 + b_1 r+b_2 r^2 + b_3 r^3+ b_4 r^4 +\mathcal{O}(r^5) \,,
\label{}
\end{equation}
\begin{equation}
\Phi = c_0 +c_1 r + c_2 r^2 + c_3 r^3 + c_4 r^4 +\mathcal{O}(r^5) \,.
\label{}
\end{equation}
The independent term $c_{0}$ in the expansion of $ \Phi$ can be set equal to zero by a transformation in the time coordinate $t$.

On the other hand, the non-zero components of the Riemann tensor are
\begin{multline}
      R_1 \equiv R^{01}{}_{01} = \frac{(2 m-r) \Phi ''}{r}\\
      +\frac{\Phi' \left[r m'+r (2 m-r) \Phi '-m\right]}{r^2} \,,
\end{multline}
\begin{equation}
R_2 \equiv  R^{02}{}_{02} = R^{03}{}_{03}= \frac{(2 m-r) \Phi '}{r^2} \,,
\end{equation}
\begin{equation}
R_3 \equiv  R^{12}{}_{12} = R^{13}{}_{13}= \frac{r m'-m}{r^3} \,,
\end{equation}
\begin{equation}
R_4 \equiv  R^{23}{}_{23} = \frac{2 m}{r^3} \,.
\end{equation}
All the algebraic  curvature invariants can be constructed from the previous expressions \cite{2008PhRvD..78f4049B}. Therefore, if all of \textcolor{black}{the latter} do not \textcolor{black} {exhibit} divergences, the spacetime is regular. The components of the Riemann tensor at the origin, \textcolor{black}{having replaced the series expansions}, are given by
\begin{align}
    R_1 &= \frac{2 a_0 \left(c_1^2+c_2\right)}{r}-\frac{a_0 c_1}{r^2}  + \mathcal{O}(r^0) \,,\\
    R_2 &= \frac{2 a_0 c_1}{r^2}+\frac{\left(2 a_1-1\right) c_1+4 a_0 c_2}{r} + \mathcal{O}(r^0) \,,\\
    R_3 &= \frac{a_2}{r}-\frac{a_0}{r^3} + \mathcal{O}(r^0) \,, \\
    R_4 &= \frac{2 a_0}{r^3}+\frac{2 a_1}{r^2}+\frac{2 a_2}{r} + \mathcal{O}(r^0) \,.
\end{align}
It is evident that the spacetime is regular at $r=0$ if $a_0 = a_1 = a_2 = c_1 = 0$. 

The energy density in Eq. \eqref{eqn:rho} around the origin is given by
\begin{align}
\begin{aligned}
8\pi\rho_{\rm effec} =& \frac{2[(680\alpha_1-3\alpha_3-20\chi_6)(1+b_0)^4]}{5r^6}\\
&+\mathcal{O}\Big(\frac{1}{r^5}\Big) \,,
\label{}
\end{aligned}
\end{align}
thus, if we do not fine tune the model parameters, we must set $b_0 = -1$ which leads to 
\begin{align}
\begin{aligned}
8\pi\rho_{\rm effec}=& \frac{2b_1^2[9-(-8\alpha_1+2\alpha_3+3\chi_5+9\chi_6)b_1^2]}{3r^2}\\
&+\frac{240b_1b_2-60b_1^3}{15r}\\
&-\frac{4b_1^3(-950\alpha_1+68\alpha_3+45\chi_5+75\chi_6)b_2}{15r}\\
&+12b_2^2 +\mathcal{O}(r^1) \,.
\label{}
\end{aligned}
\end{align}
Finally, we set $b_1=0$  in order to remove the remaining singularities in the energy density. Taking into account the previous conditions on the coefficients, we have solved the field equations 
and found that the remaining coefficients are given by
\begin{equation}
a_3=2b_2^2 \,,
\label{}
\end{equation}
\begin{equation}
a_5=\frac{3}{5}\mu^2b_2^2-\frac{8b_2^3}{5}+\frac{172 \alpha_1 b_2^4}{3}+\frac{7 \alpha_3 b_2^4}{15}-4\chi_6 b_2^4 \,,
\label{}
\end{equation}
\begin{multline}
b_4=\frac{\mu^2b_2}{10}-\frac{3b_2^2}{10}+\frac{4b_2^3}{5}+\alpha_1 b_2^3\\
+\frac{7 \alpha_3b_2^3}{10}
+\frac{\chi_5 b_2^3}{5}-\chi_6 b_2^3 \,,
\end{multline}
\begin{equation}
c_2=2 b_2^2 \,,
\label{}
\end{equation}
\begin{multline}
c_4=\frac{\mu^2b_2^2}{5}-\frac{4b_2^3}{5}+\frac{12b_2^4}{5}-8\alpha_1b_2^4\\
+\frac{9\alpha_3 b_2^4}{10}-\frac{2\chi_5b_2^4}{5}-2\chi_6b_2^4 \,,
\label{}
\end{multline}
\begin{equation}
a_4=a_6=0 \,,    
\end{equation}
\begin{equation}
b_3=b_5=0 \,,    
\end{equation}
and
\begin{equation}
c_3 = c_5=0 \,.    
\end{equation}
The coefficient $b_2$ is arbitrary and has been taken as the shooting parameter in the subsequent numerical analysis. The parameter $\chi_4$ has no contribution in the Euler-Lagrange 
equations \textcolor{black}{because of the implicit symmetry between the background metric and the  vector field configuration that leads to a vanishing reduced Lagrangian\footnote{$\chi_4$ does not contribute either in cosmological applications \cite{Garnica:2021fuu}.}.} We have also found that the effective energy density at $r=0$, given by the non-vanishing first term of the expansion, is given by
\begin{equation}
\rho_{\rm effec}\Big|_{r=0} =\frac{3b_2^2}{2 \pi} \,,
\label{eqn:rho0}
\end{equation}
\textcolor{black}{which implies the  shooting parameter is related to the central density of the object}.  The asymptotic solutions not only exhibit the behaviour of the fields near the origin but also serve to start the numerical integration from a point $r\approx0$, given that the field equations are  singular at the origin. 

We can see that the leading contributions \textcolor{black}{in all expansion series} come from the minimally coupled EYM theory. However, this does not mean the new terms produce only small changes both in the \textcolor{black}{local and global} behaviour \textcolor{black}{of the solutions}. As we will see later, in the region $r\approx 10 - 10^2$ where the vector field has a nontrivial behaviour $w\neq 1$ (transition zone), there are appreciable changes \textcolor{black}{in the space-time structure  and the vector field profile}.  \textcolor{black}{These changes are} induced by the small differences near the origin coming from the new terms of the GSU2P theory. 

\subsection{Asymptotic series solution at spatial infinity}
We have proceeded to study the model at spatial infinity and first we have performed an asymptotic expansion in inverse powers of $r$ as follows: 
\begin{equation}
m = M+\frac{\tilde{a}_1}{r}+\frac{\tilde{a}_2}{r^2}+\frac{\tilde{a}_3}{r^3}+\mathcal{O}\left(\frac{1}{r^4}\right) \,,
\label{eqn:m-asymp-infty}
\end{equation}
\begin{equation}
w = w_\infty+\frac{\tilde{b}_1}{r}+\frac{\tilde{b}_2}{r^2}+\frac{\tilde{b}_3}{r^3}+\mathcal{O}\left(\frac{1}{r^4}\right) \,,
\label{eqn:w-asymp-infty}
\end{equation}
\begin{equation}
\Phi = \Phi_\infty+\frac{\tilde{c}_1}{r}+\frac{\tilde{c}_2}{r^2}+\frac{\tilde{c}_3}{r^3}+\mathcal{O}\left(\frac{1}{r^4}\right) \,.
\label{eqn:phi-asymp-infty}
\end{equation}
The Proca Lagrangian piece (the one that involves the vector field mass $\mu$) induces an exponential decay that is dominant with respect to the other Lagrangian pieces of the theory so that the resulting solution is the same as that found in Ref. \cite{1993PhRvD..47.2242G}. That is the reason why such a term was not taken into account in the series\footnote{In addition to the fact that the reasoning behind the construction of the GSU2P is the breaking of the internal gauge invariance. This is possible to attain in more ways than just adding a mass term.}.

The value of the gravitational mass $M$ is determined after solving numerically the boundary value problem, i.e., after finding the value of the shooting parameter $b_2$ so that, in the limit $r\to \infty$, the functions $m\to M=\mbox{const.}$, $\Phi\to\mbox{const}$, and $w\to w_{\infty}=\mbox{const.}$ The value of $\Phi_{\infty}$ is also determined after performing the numerical integration. The latter can be renormalized to 0, with the appropriate coordinate transformation in the variable $t$, so that the spacetime is asymptotically flat. We have found three different kinds of asymptotic behaviour \textcolor{black}{for the involved variables} at infinity classified according to the value of $w_{\infty}$. The \textcolor{black}{corresponding} values of the asymptotic coefficients can be found in the Tab. \ref{tab:coeff_infty}. \textcolor{black}{The main physical properties of the found asymptotic solutions will be discussed below}.

\begin{table*}[t]
\begin{ruledtabular}
\begin{tabular}{ r |r | c | c | c}
& & $w_\infty = -1$ & $w_\infty = 1$ & $w_\infty = w_{\rm I,II}$\\
\hline
\multirow{5}{*}{$m$}& $\tilde{a}_1$ & 0 & 0 & $-Q^2_{\rm I,II}/2$\\
&$\tilde{a}_2$ & 0 & 0& 0\\
&\multirow{3}{*}{$\tilde{a}_3$} & \multirow{3}{*}{$-\tilde{b}_1^2$} &\multirow{3}{*}{$64(25\alpha_1 - 36 \alpha_3 + 5\chi_6)/15 -\tilde{b}_1^2$} & $\frac{2}{3} \tilde{b}_2 (w_\infty+1) [\chi_{12}$\\
& & & & $+(\chi_{12}-1) w_\infty^2+2 \chi_{12} w_\infty +w_\infty]$\\
& & & & $+\frac{4}{15} (w_\infty+1)^4 (25\alpha_1-36 \alpha_3+5 \chi_6)$\\
\hline
\multirow{5}{*}{$w$}& $\tilde{b}_1$ & free & free & 0\\
&$\tilde{b}_2$ &  $3(2 M - \tilde{b}_1)\tilde{b}_1/4$& $16 \alpha_1 + 6 \alpha_3 - 16\chi_6 + 3(2 M + \tilde{b}_1)\tilde{b}_1/4$ &$ \frac{(w_\infty+1)^3 (8 \alpha_1+3 \alpha_3-8 \chi_6)}{3 \chi_{12} (w_\infty+1)^2 -3 w_\infty^2 +7}$\\
&\multirow{3}{*}{$\tilde{b}_3$} & \multirow{3}{*}{$\tilde{b}_1[48 M^2 - 42M\tilde{b}_1 + (11 - 2\chi_{12})\tilde{b}_1^2]/20$}& $(512\alpha_1 M + 96\alpha_3 - 512\chi_6 M)/20$ &\multirow{3}{*}{$\frac{2 M \left(8 \tilde{b}_2-3 \alpha_3 (w_\infty+1)^3\right)}{3 \chi_{12} (w_\infty+1)^2 -3 w_\infty^2 +13}$}\\
& & & $+ 16\tilde{b}_1(26\alpha_1 + 13\alpha_3 + 2 \chi_5 - 26\chi_6)/20$&\\
& & & $+(48 M^2 - 42M\tilde{b}_1 + 11\tilde{b}_1^2)/20$ &\\
\hline
\multirow{3}{*}{$\Phi$}&$\tilde{c}_1$ & $-M$ & $-M$ & $-M$\\
&$\tilde{c}_2$ & $-M^2$ & $-M^2$ & $-M^2+Q_{\rm I,II}^2/2$\\
& $\tilde{c}_3$ & $-4M^3/3$ &  $-4M^3/3$ & $-4M^2/3 + Q_{\rm I,II}^2 M$ \\
\end{tabular}
\caption{\textcolor{black}{Values of the coefficients of the asymptotic expansion in inverse powers of $r$ with integer exponents, see Eqs. \eqref{eqn:m-asymp-infty}-\eqref{eqn:phi-asymp-infty}\label{tab:coeff_infty}. The solutions are classified according to the asymptotic value of $w$ denoted by $w_{\infty}$. Globally neutral solutions  ($\tilde{a}_1=0$) are characterized by $w_{\infty}=\pm1$ (first and second columns); the coefficient $\tilde{b}_1$ is free for these cases. Globally charged solutions ($\tilde{a}_1 \neq 0$) are characterized by $w_{\infty}=w_{\rm I,II}$ (third column) which correspond to the values of the vector field found in the analytic Reissner-Nordstr\"om solution, see Eq. \eqref{eqn:sol_wc2};  also in this case, the charges $Q_{\rm I,II}$ correspond to the values given by Eq. \eqref{eqn:charge-RN-sol}.}}
\end{ruledtabular}
\end{table*}
\subsubsection{Globally neutral solutions}
The first kind of solution is characterized by the absence of an effective global charge and by the fact that the vector field approximates in the limit to the asymptotic values $w\to \mp1$. The absence of a global charge is a consequence of the faster decay of the mass function, i.e. $\tilde{a}_1=\tilde{a}_2 = 0$. When $w
\to-1$, the asymptotic behaviour is practically indistinguishable from the EYM case. Nevertheless, there are subdominant contributions coming explicity from the GSU2P theory (see, for instance, the coefficient $\tilde{b}_3$ in Tab. \ref{tab:coeff_infty}). The  equation of state parameter  $w_{\rm eff}$ is given by

\begin{align}
\begin{aligned}
w_{\rm eff}=&\frac{1}{3}+\frac{[6250 \alpha_1-6133\alpha_3-30(\chi_5+3\chi_6)]\tilde{b}_1^{2}}{135r^{4}}\\
&+\mathcal{O}\left(\frac{1}{r^5}\right) \,,
\label{}
\end{aligned}
\end{align}
thus the vector field behaves asymptotically as radiation.

The GSU2P theory breaks the reflection symmetry $w\mapsto -w$ and gives rise to a new kind of solution with $w\to 1$. Again, these solutions do not exhibit a global charge, but the decay of the vector field has new contributions coming directly from the GSU2P terms. In this case, the equation of state parameter does not approximate in the limit exactly to $1/3$ due to the correction arising from the model parameters:
\begin{equation}
    \omega_{\rm eff} = \frac{1}{3} - \frac{8320 \alpha_1-7152 \alpha_3-640 \chi_6}{192 (25 \alpha_1-36 \alpha_3+5 \chi_6)-45 \tilde{b}_1^2} + \mathcal{O}\biggl(\frac{1}{r}\biggr) \,.
    \label{eqn:omegap1}
\end{equation}
Although this result \textcolor{black}{reveals} that the vector field does not behave \textcolor{black}{exactly} as radiation \textcolor{black}{at} infinity, we have found numerically that the correction is negligible \textcolor{black}{for asymptotically flat solutions} (see subsection. \ref{ssec:nonminimal}). 
\subsubsection{Globally charged solutions} \label{sec:globalcharge}
Globally charged solutions, i.e., $\tilde{a}_1 \neq 0$, constitute a special case \textcolor{black}{that demands}  \textcolor{black}{a non-vanishing} $\chi_{12}$. These solutions approximate in the limit to the exact solution found in Section \ref{sec:analytical}. \textcolor{black}{We have obtained two solutions at spatial infinity as a post-Reissner-Nordstr\"om expansion in inverse powers of $r$, one with non-integer exponents and another with integer exponents (for the latter, see the third column in Tab. \ref{tab:coeff_infty}).} The charged solution with non-integer exponents is given by
\begin{equation}
 m = M - \frac{Q^2_{\rm I}}{2 r} - \frac{d^2D}{r^{2\beta+1}} + \mathcal{O}\biggl(\frac{1}{r^{3\beta+1}}+\frac{1}{r^{2\beta+2}}\biggr) \,,  \label{eqn:m-charged-asymp}
\end{equation}
\begin{equation}
    w = w_{\rm I} + \frac{d}{r^\beta}  +\mathcal{O}\biggl(\frac{1}{r^{\beta+1}}+\frac{1}{r^{2\beta}}\biggr) \,, \label{eqn:w-charged-asymp}
\end{equation}
\begin{equation}
    \delta = -\frac{d^2 \beta^2}{(1+\beta)}\frac{1}{r^{2\beta+2}}+\mathcal{O}\biggl(\frac{1}{r^{2\beta+3}}+\frac{1}{r^{3\beta+2}}\biggr) \,, \label{eqn:delta-charged-asymp}
\end{equation}
where $d$ is \textcolor{black}{also a} free parameter and
\begin{equation}
    \beta = -\frac{1}{2}\Biggl(1-\sqrt{\frac{3+15\chi_{12}+6\sqrt{1+8\chi_{12}}}{1-\chi_{12}}}\Biggr) \,,
\end{equation}
\begin{equation}
D = \frac{1+8\chi_{12}+ \sqrt{1+8\chi_{12}} - 2\beta^2(\chi_{12}-1) }{2(1+2\beta)(\chi_{12}-1)} \,,
\end{equation}
\textcolor{black}{are quantities fully determined once $\chi_{12} $ is specified}. \textcolor{black}{Solutions approximating in the limit to the  Reissner-Nordstr\"om II solution as an expansion in inverse powers of $r$ do not exist.}
The parameter $\beta$ is always positive and real if $-1/8<\chi_{12}<1$. The effective charge is imaginary when $\chi_{12}>0$ and real when $\chi_{12}\leq0$. 
\section{Numerical Solution} \label{sec:numerical}
\begin{figure*}[!t]
    \centering
    \includegraphics[width=\textwidth]{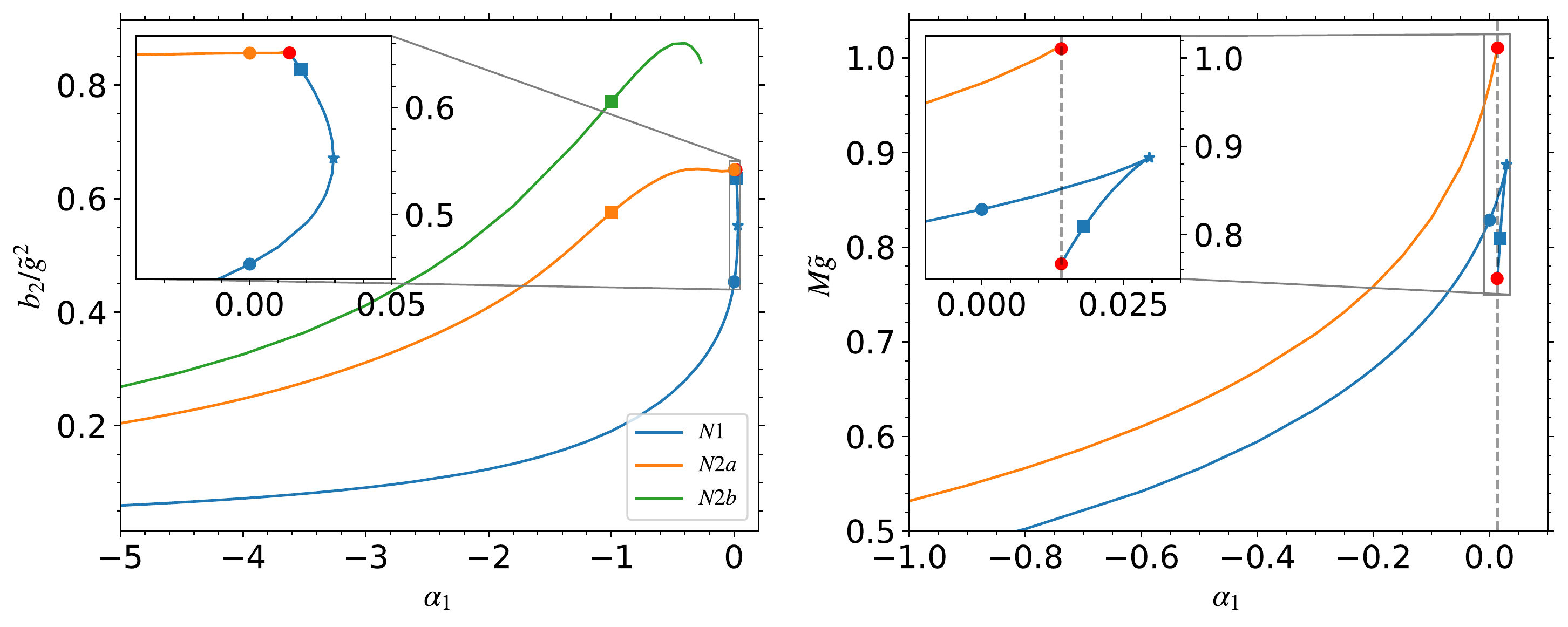}
    \caption{GSU2P particle-like solutions with $\alpha_1\neq0$. The families can be classified according to the number of zeros of the vector field $w$. We show one family \textcolor{black}{($N1$)} with one node (blue) and two families \textcolor{black}{($N2a$ and $N2b$)} with two nodes (orange and green, \textcolor{black}{respectively}). \textcolor{black}{On the left, we show the equilibrium sequences where the shooting parameter $b_2$ is a function of the parameter $\alpha_1$. On the right, we also show the equilibrium sequences but, this time, replacing the shooting parameter by the asymptotic gravitational mass.} For the one node family, there is a maximum mass, \textcolor{black}{indicated by} a star, which is a turning point \textcolor{black}{(a point where two sequences of one-node solutions merge)}. The EYM solution corresponds to the circle, and the square corresponds to a  particular solution after the turning point. Finally, there is another turning point, red circle, where the families $N1$ and $N2a$ merge in the $b_2-\alpha_1$ plane. The two-node families are more massive than the one-node family. In the plane $M-\alpha_1$ the two red circles correspond to the latter turning point mentioned above, \textcolor{black}{featuring a first-order phase transition} between the one-node to the two-node family. The orange and green squares are representative solutions of the $N2a$ and $N2b$ families (see Fig. \ref{fig:PLA1N2}). The orange circle marks the two-node EYM solution belonging to the $N2a$ family.}
    \label{fig:sequences_A1}
\end{figure*}
The field equations are singular at the origin, preventing the beginning of the numerical integration. \textcolor{black}{In order to overcome this difficulty}, we have used the series expansions at the origin to find the values of the fields and its derivatives at $r=1\times 10^{-3}$, and have used the latter as the initial conditions. We have integrated several times with different initial conditions, i.e., different values of the parameter $b_2$, until we have found a solution satisfying the boundary conditions at infinity, namely an asymptotically flat spacetime, and finite energy and stresses associated with the vector field. We have used the free parameters in the Lagrangian of Eq. \eqref{eqn:lagrangian} to construct families of solutions. 

For the sake of simplicity, we have studied each parameter of the GSU2P Lagrangian in Eq. \eqref{eqn:lagrangian} separately \textcolor{black}{which, in turn, has allowed us to understand more clearly the physical features behind each Lagrangian piece}. \textcolor{black}{When we have analyzed numerically each parameter, we have set all the other parameters to zero.} As a result, we have found various solutions that generalize the canonical EYM case which are encapsulated in \emph{one-parameter} equilibrium sequences. As usual, the solutions are classified according to the vector field zeros (nodes). 
\textcolor{black}{As already discussed, the conditions of regularity at the origin imply $w(0)=-1$. On the other hand, asymptotic flatness, in the case of globally neutral families, implies that the vector field tends to 1 or -1. Solutions whose asymptotic value is 1 inevitably exhibit an odd number of nodes. In principle, solutions that tend to -1 can be void of nodes. However, we have not found any zero-node solution, just those with an even finite number of nodes. For globally (effective) charged solutions, we have not found zero-node solutions for the values of $\chi_1$ and $\chi_2$  we have considered.} 

Some solutions exhibit new phenomena such as \textcolor{black}{zones with} negative energy density, imaginary (effective) charge, and global charge.  \textcolor{black}{All the solutions we have found have \emph{finite and positive} total gravitational-mass, even in the cases with regions of negative energy density.}

The mass parameter $\mu$ was not included in this work since it was already analyzed in Ref. \cite{1993PhRvD..47.2242G} and because we have wanted to explore the pure effects of the GSU2P theory. There, it was shown that the mass of the vector fields has an upper limit of $4.454 \times 10^{-2} \tilde{g}^2$ to generate asymptotically flat  solutions.

\textcolor{black}{The numerical solutions show that the behaviour of the effective charge can be used qualitatively to discriminate between three different regions: 1) a high-density interior region with approximately zero charge, 2) a transition and near-field region where the metric is approximately Reissner-Nordstr\"om, and 3) a far-field region where the metric approximates to either the Schwarzschild's solution as $r \rightarrow \infty$, in the case of globally neutral solutions, or to the Reissner-Nordstr\"om solution, in the case of globally charged solutions.}
\subsection{Non-minimal coupling}\label{ssec:nonminimal}

Non-minimal couplings of the metric to the vector field correspond to the  cases $\alpha_1\neq0$ and $\alpha_3\neq0$. These non-minimal couplings introduce new properties such as the merging of the one- and  two-node families in \textcolor{black}{comparison to the minimal coupling EYM case.}

The first case we have analyzed corresponds to $\alpha_1\neq0$. We have found three different families of solutions: one family with one node and two families with two nodes, which we have called $N2a$ and $N2b$, respectively. These families are shown in Fig. \ref{fig:sequences_A1}. Other families of solutions with more zeros can be found but, for our present purposes, these three families are sufficient.

The GSU2P theory introduces new effects in the behaviour of soliton solutions. \textcolor{black}{In particular,}
the $\alpha_1$ one-node family presents regions with imaginary charge and negative energy density. As can be seen in Fig. \ref{fig:sequences_A1}, \textcolor{black}{the $\alpha$ parameter has a maximum value of $0.0295$ (indicated by a blue star), beyond which no one-node asymptocally flat solutions have been found. } Also, from the left plot in Fig. \ref{fig:sequences_A1}  and from Eq. \eqref{eqn:rho0}, we have deduced that the central effective energy density decreases as $\alpha_1$ becomes more negative. 

The gravitational mass has its maximum at $\alpha_1= 0.0295$, indicated by the star in the right plot of Fig \ref{fig:sequences_A1}. This place  \textcolor{black}{represents a \emph{turning point} in the equilibrium sequence}, i.e, two sequences of one-node solutions merge at this point. This feature is better displayed in the zoomed regions of Fig \ref{fig:sequences_A1}. There is another turning point located at $(b_2 \approx 0.651, \alpha_1 \approx 0.0137)$, where the family \textcolor{black}{$N_1$} merges with the two-node family $N2a$ (see the red circles in Fig. \ref{fig:sequences_A1}). At this point, the physical character of the solutions changes dramatically, as is revealed by the discontinuous change in the gravitational mass shown in the right plot of Fig. \ref{fig:sequences_A1}.
\begin{figure}[!b]
    \centering
    \includegraphics[width=0.47
    \textwidth]{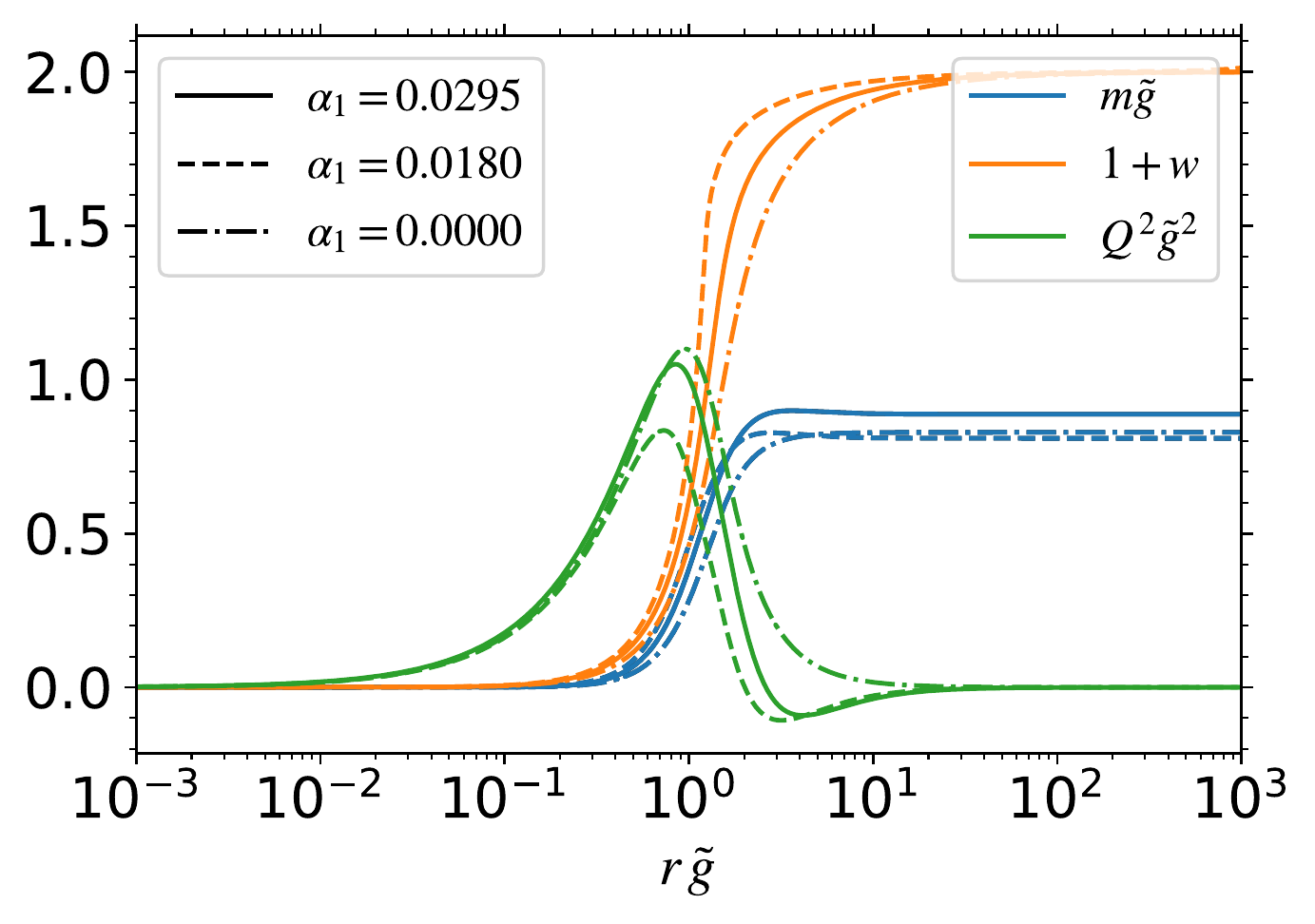}
    \caption{Comparison and contrast of representative GSU2P solutions belonging to the one node family with respect to the EYM solution. The blue curves correspond to the mass function, the orange curves correspond to the vector field $w$, and the green ones correspond to the effective charge. For all the field variables, the dotted-dashed, the continuous, and the dashed curves correspond to the EYM case (the circle in Fig. \ref{fig:sequences_A1}), the maximum (the star in Fig. \ref{fig:sequences_A1}), and the solution after the maximum (the square in Fig. \ref{fig:sequences_A1}), respectively. The solutions near and after the maximum are characterized by the presence of a local imaginary effective charge and negative energy densities.}
    \label{fig:PLA1}
\end{figure}
\begin{figure}[!t]
    \centering
    \includegraphics[width=0.47
    \textwidth]{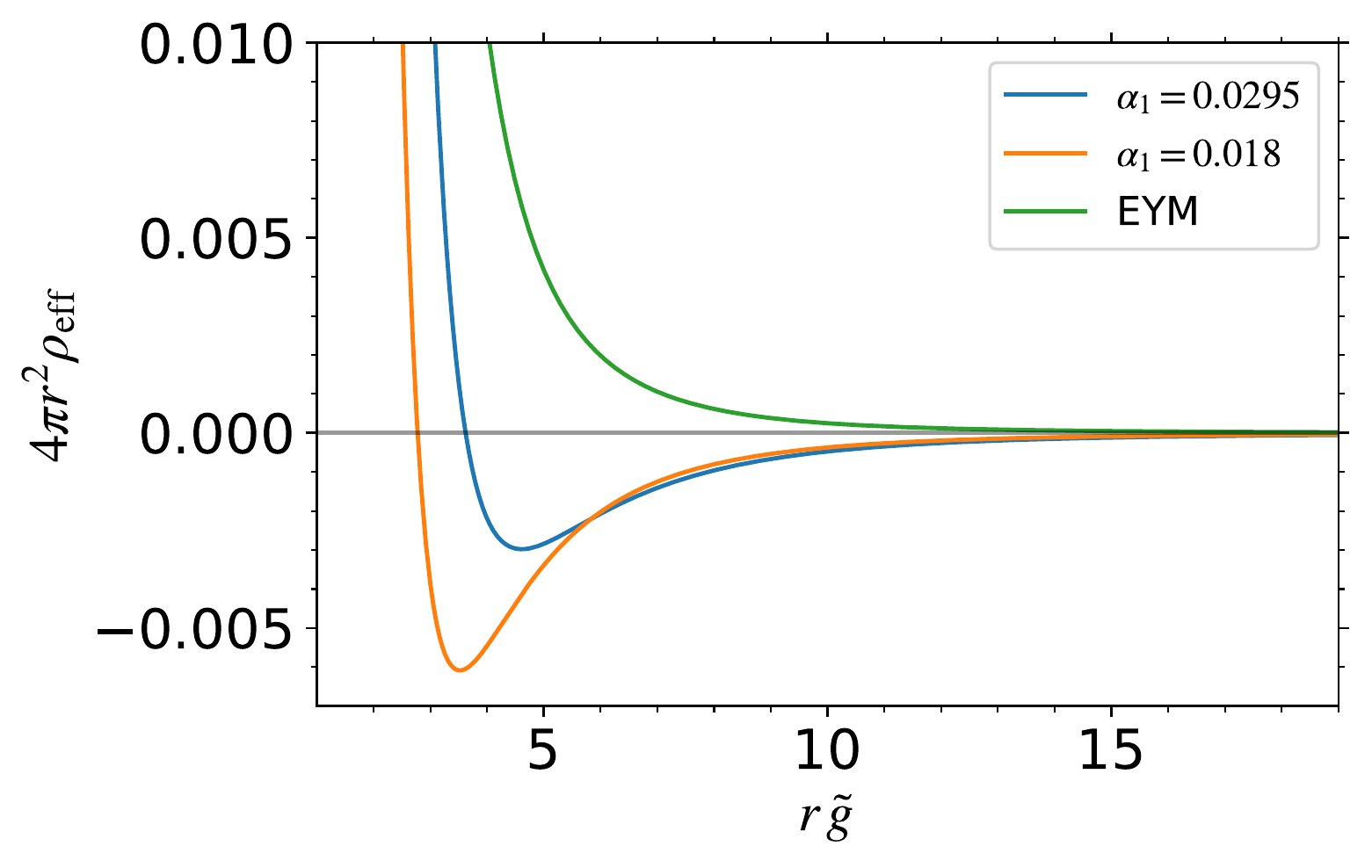}
    \caption{Effective energy density of the $\alpha_1$ one-node family for selected values of the parameter as indicated in the legend. The energy density is negative in the transition region when the parameter is near the turning point. This plot shows the cases $\alpha_1 =0.0295$ (blue) $\alpha_1=0.018$ (orange) and $\alpha_1=0$, EYM (green).}
    \label{fig:rhoEffA1N1}
\end{figure}
\begin{figure}[!b]
    \centering
    \includegraphics[width=0.5\textwidth]{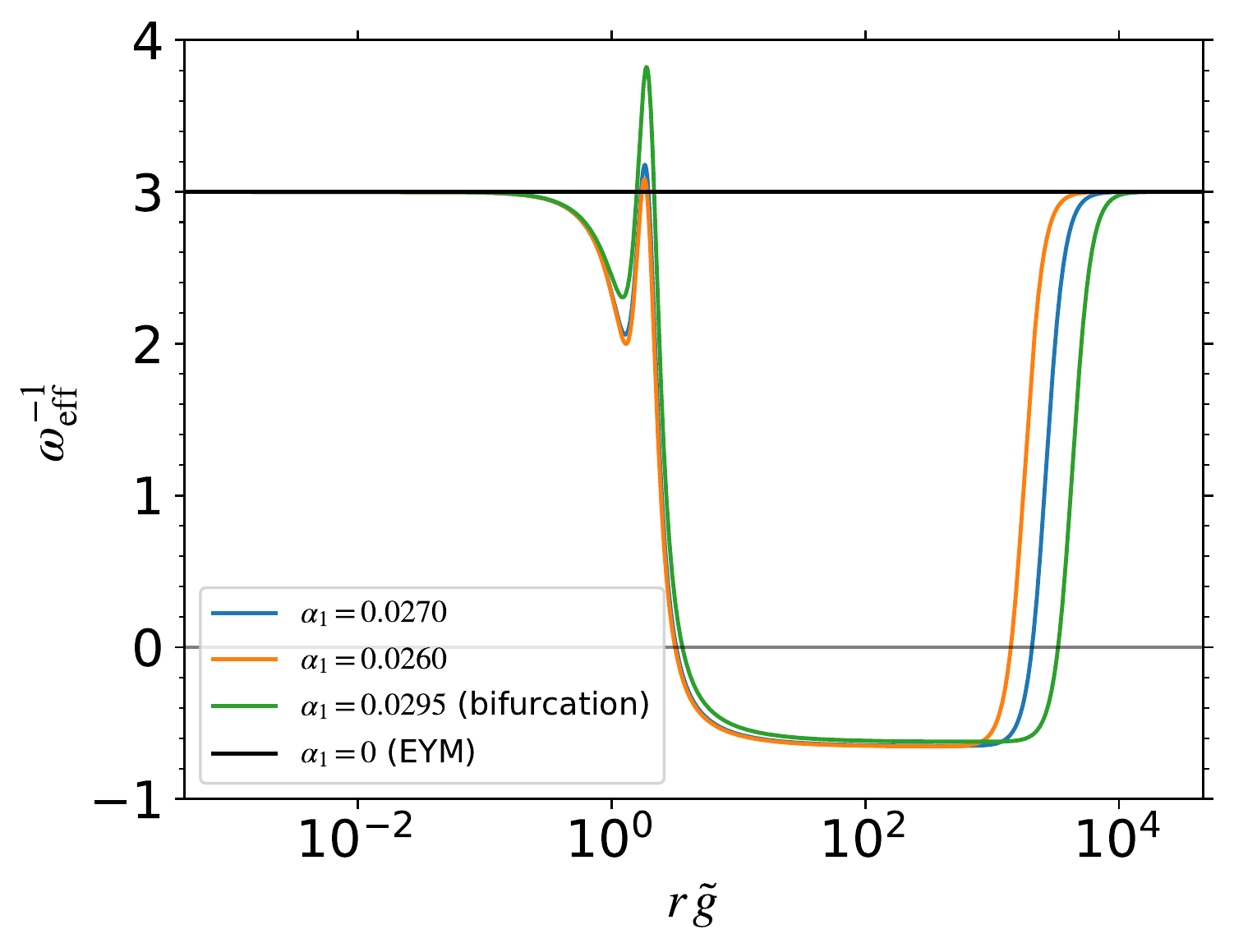}
    \caption{Inverse of the effective parameter of state, $\omega_{\rm eff}^{-1}\equiv \rho_{\rm eff}/p_{\rm eff}$, for some of the \textcolor{black}{\emph{potentially stable}} solutions of the $\alpha_1$  one-node family. It can be seen that, near the origin and at infinity, the vector fields behave as radiation. \textcolor{black}{$\omega_{\rm eff}^{-1}$ is depicted here since the energy density becomes null and then negative in some regions.}}
    \label{fig:omega_alpha1_N1}
\end{figure}


To appreciate more features of this family, we show three representative solutions in Fig. \ref{fig:PLA1}: one at the location of the maximum mass $\alpha_1=0.0295$, another one before the latter point at $\alpha_1 = 0$, and the last one at $\alpha_1 = 0.018$, near the point where the family merges with the two-node family. \textcolor{black}{The solutions near the EYM solution ($\alpha_1 = 0$) exhibit a positive energy density  and real effective charge \textcolor{black}{everywhere}. In contrast, solutions near the maximum mass have \emph{regions} where the energy density is negative  and the effective charge is imaginary, see Fig. \ref{fig:rhoEffA1N1}}; the same behaviour is encountered for the whole equilibrium sequence that contains the blue square (see the right plot in Fig. \ref{fig:sequences_A1}). 
\textcolor{black}{The latter two properties and the existence of turning points are not present in the canonical EYM case, {\it so they represent novel features of the GSU2P theory}. As will be discussed later, {\it the existence of turning points could indicate places where the stability of the solutions changes favorably from unstable to stable solutions}}.

Regarding the effective equation of state parameter $\omega_{\rm eff}$, we have found numerically that the vector fields behave as radiation in the limit $r\to \infty$, see Fig. \ref{fig:omega_alpha1_N1}, i.e., the correction to $1/3$ given by Eq. \eqref{eqn:omegap1} is negligible.

\begin{figure}[!t]
    \centering
    \includegraphics[width=0.47\textwidth]{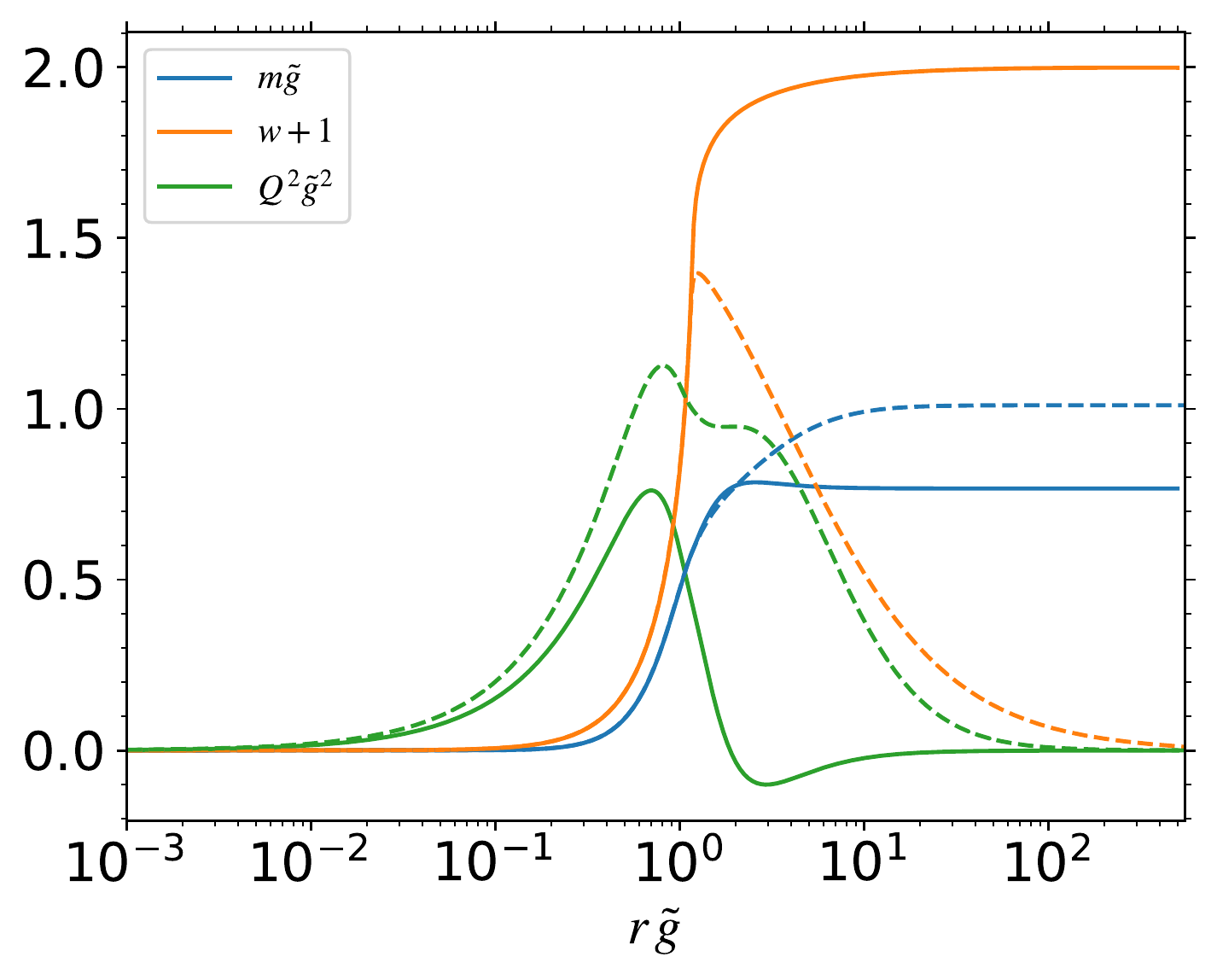}
    \caption{\textcolor{black}{Comparison of the field variables near the merging point of the one-node (continuous curves) and two-node (dashed curves) families for $\alpha_1\approx 0.137$. The value of the asymptotic mass changes discontinuously.}}
    \label{fig:phasetrans}
\end{figure}

\begin{figure}[!b]
    \centering
    \includegraphics[width=0.5\textwidth]{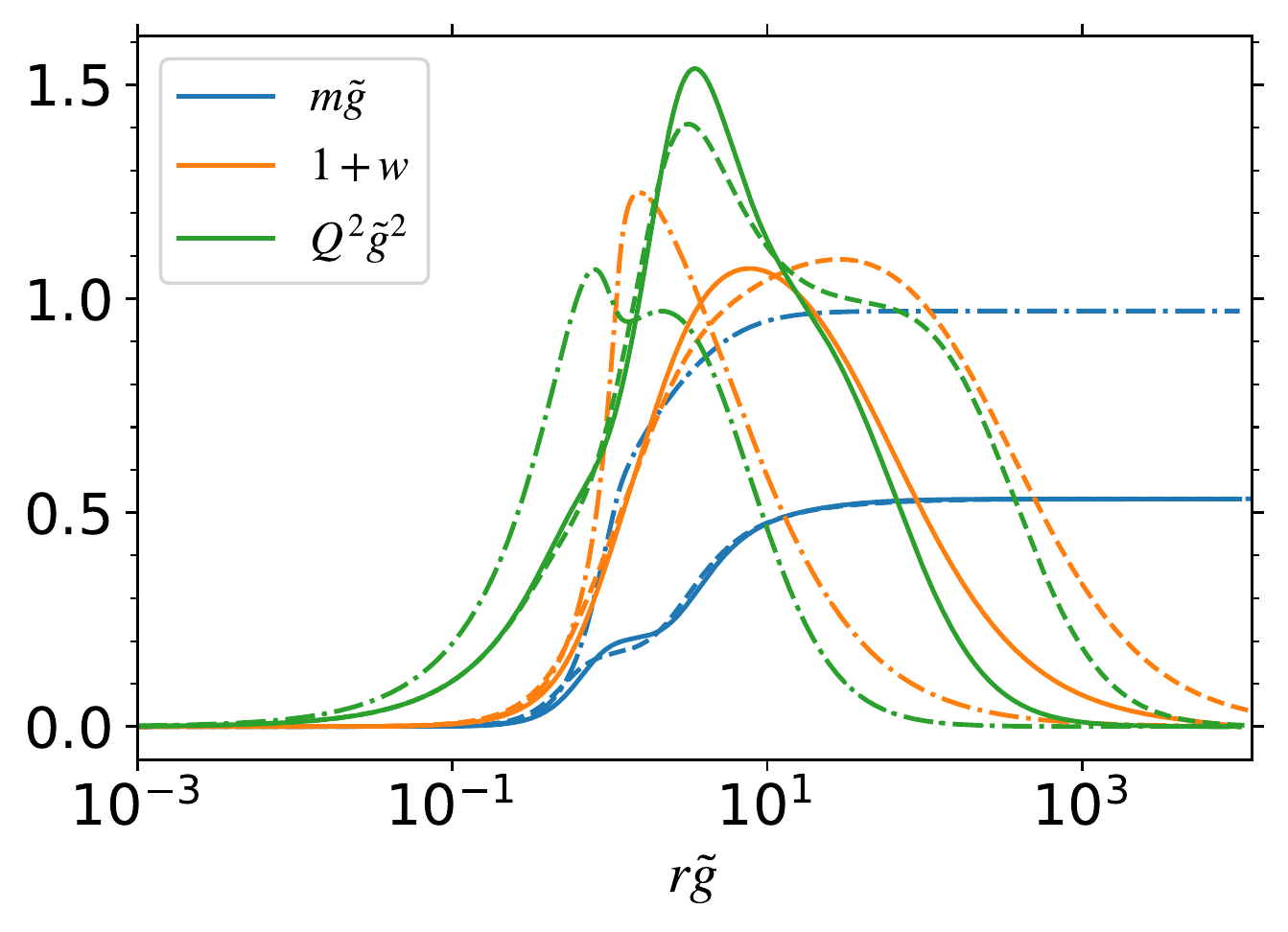}
    \caption{Representative solutions of the $N2a$ and $N2b$ families, both with $\alpha_1 =-1$, corresponding to the orange and green squares in the Fig. \ref{fig:sequences_A1}, respectively. The continuous curves correspond to the $N2a$ family, the dashed curves correspond to the $N2b$ family, and the dotted-dashed curves correspond to the EYM case, i.e., $\alpha_1=0$. }
    \label{fig:PLA1N2}
\end{figure}

For the $N2a$ family, the mass is a monotonically increasing function of the parameter $\alpha_1$ up to the point $b_2 \approx 0.651, \alpha_1 \approx 0.0137$ where the family merges \textcolor{black}{with the one-node family in the $b_2-\alpha_1$ plane. The nature of the solutions changes drastically at this point, which can be seen in the discontinuous change in the total gravitational mass. The vector field profiles belonging to one-node and two-node families cannot be continuously deformed into each other because of the different boundary conditions at infinity\footnote{\textcolor{black}{These two families are not homotopic.}} (see Fig. \ref{fig:phasetrans}).}  In agreement with Fig. \ref{fig:PLA1N2}, for $\alpha_1<0$, the vector field and charge decay more slowly than in the EYM case; besides, the gravitational mass is smaller than in the EYM case. For $\alpha_1 \lesssim-2$, there exist regions with negative energy density.

Regarding the $N2b$ family, we have found $\alpha_1 <0$ as a necessary condition to get asymptotically flat solutions. According to Fig. \ref{fig:PLA1N2}, we have also found that the gravitational mass is always smaller than in the EYM case. The vector field decays even more slowly than in the case of the previous two-node family, and the charge has a larger extension.

\begin{figure*}[!t]
    \centering
    \includegraphics[width=
    \textwidth]{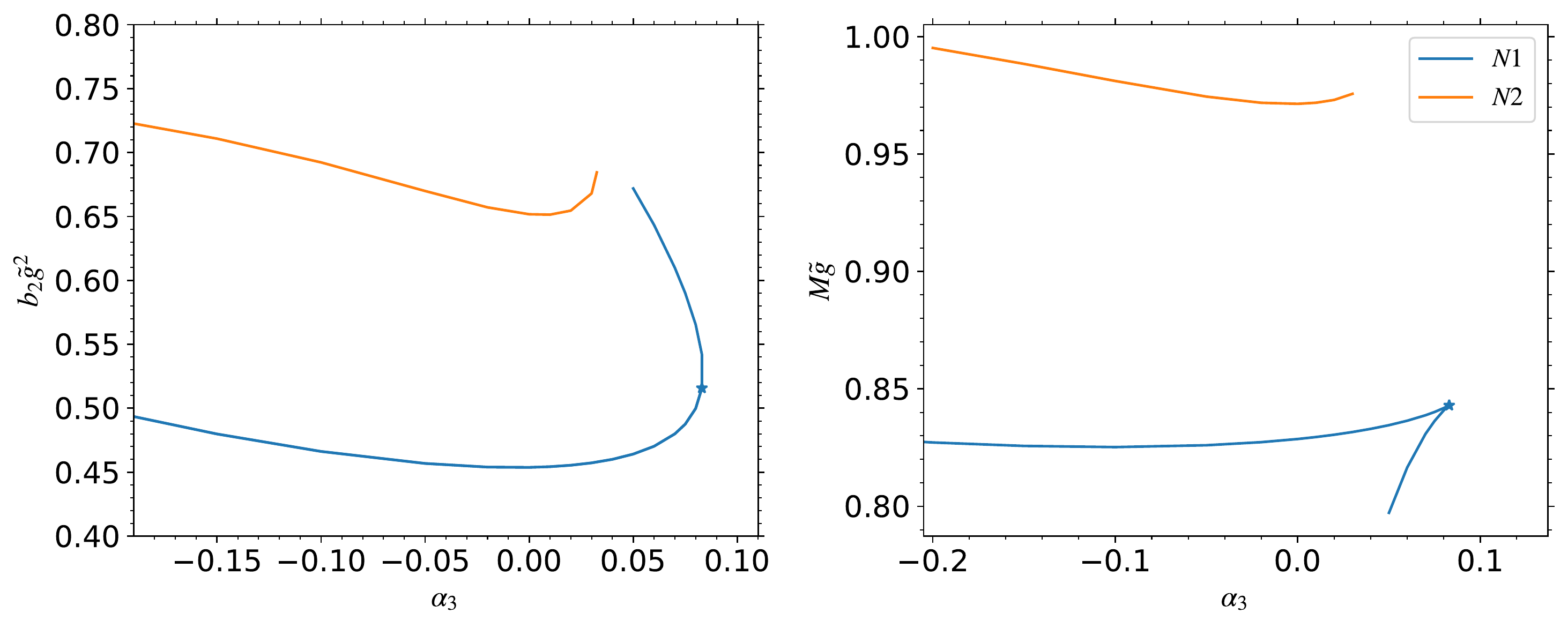}
    \caption{Family of GSU2P particle-like solutions parametrized by $\alpha_3$. We show the one-node family (blue) and the two-node family (orange). For the one-node family, there exists a turning point labeled here with a star. }
    \label{fig:my_label}
\end{figure*}

\textcolor{black}{The case $\alpha_3$, see Fig. (\ref{fig:my_label}), exhibits similar behaviour to the previous one in the region near the EYM solution. It shares the property of having a turning point in the one-node family. However, given the accuracy of the numerical methods used to integrate the equations, we have not been able to find a bifurcation of the one-node family to the two-node family. Moreover, the mass and the central density grow  as $\alpha_3$ becomes more negative. In contrast to the $\alpha_1$ case, we have not found a maximum mass for the families we have studied.}
\subsection{Minimal coupling}
\subsubsection{Case $\chi_1, \chi_2$}
This case is of special interest since new globally charged solutions have been found. The free parameters $\chi_1$ and $\chi_2$ can be absorbed in the field equations into a single parameter $\chi_{12}\equiv2\chi_1 + \chi_2$. From the analytical solution studied in Sec. \textcolor{black}{\ref{sec:analytical}}, we expect to obtain two families of solutions that approximate in the limit to the I and II Reissner-Nordstr\"om cases. However, as has been shown in Sec. \ref{sec:globalcharge}, only the solution I can be approached as inverse powers of $r$ in the limit $r\to \infty$. 

{\it Interestingly, there are no globally charged particle-like solutions in the EYM or the massive case. Globally charged solutions in the GSU2P theory arise only when new quartic terms in the vector field are considered} \textcolor{black}{as we shall see}. The mass term prevents an asymptotic solution of a Reissner-Nordstr\"om type. In contrast, the \textcolor{black}{pure} Yang-Mills self-interaction term is responsible for the existence of the analytical EYM globally charged solution. The asymptotic solution in Eqs. \eqref{eqn:m-charged-asymp}-\eqref{eqn:delta-charged-asymp} implies that $\chi_{12} \neq0$ is necessary to obtain an approximately charged solution; otherwise, the only charged solution is the trivial EYM Reissner-Nordstr\"om (non-regular) solution. 

\begin{figure}
    \centering\includegraphics[width=0.49\textwidth]{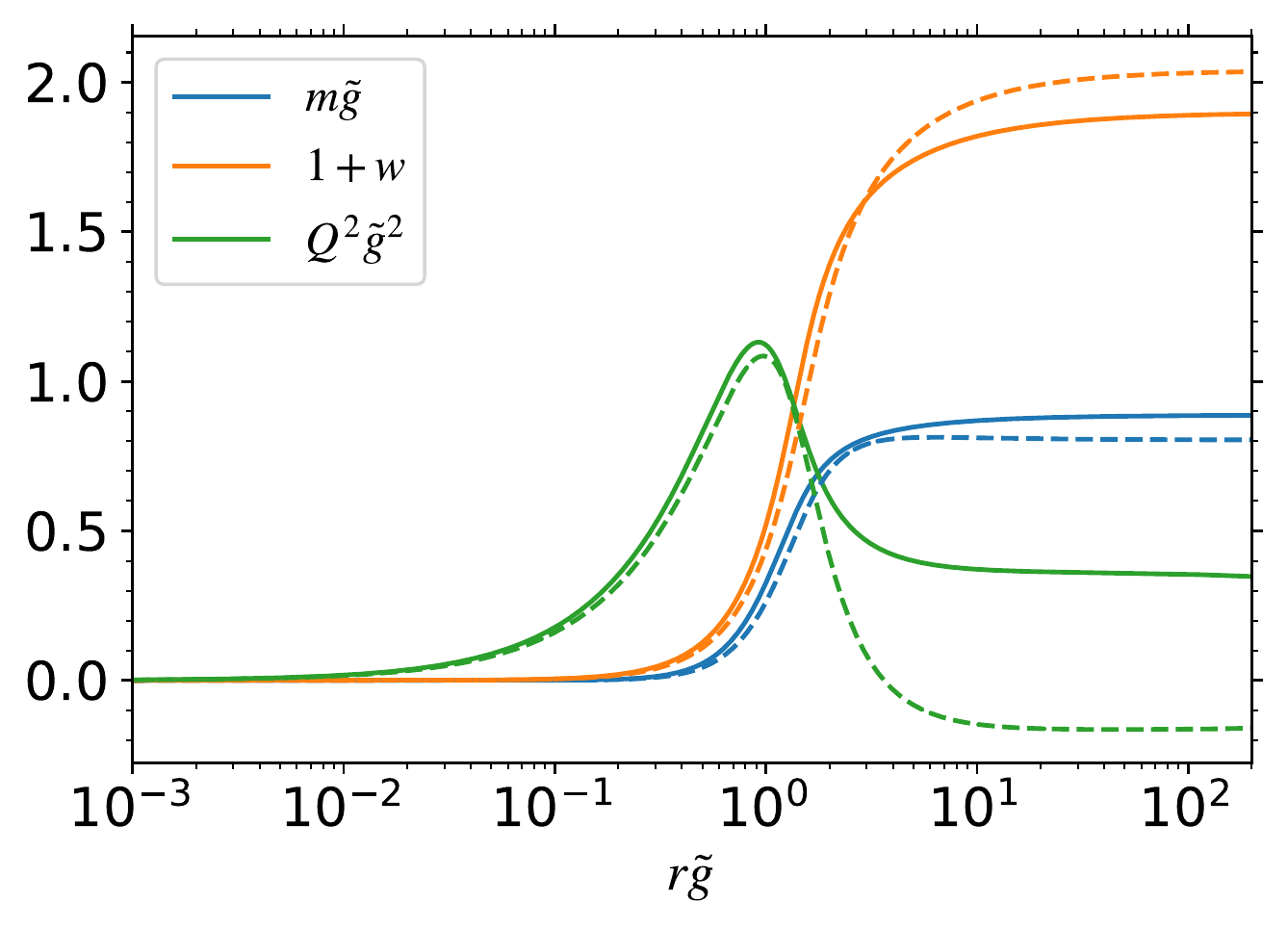}
    \caption{Mass function (blue), vector field (orange) and effective charge (green) of two representative members of the globally charged solutions of the  GSU2P $\chi_{12}$ family. The continuous curves correspond to the case $\chi_{12} = 0.01 $ with \emph{real} charge. The dashed curves correspond to the case $\chi_{12}= -0.025$ with \emph{imaginary} charge.}
    \label{fig:PLC12N1}
\end{figure}

Solutions approximating asymptotically to the Reissner-Nordstr\"om I space-time  \textcolor{black}{geometry} have been obtained numerically, and the different kinds of behaviour of the field variables for representative values of $\chi_{12}$ are shown in Fig. \ref{fig:PLC12N1}. {\it \textcolor{black}{We have been able to conclude that the new non-derivative self-interactions present in the GSU2P theory give way to the creation of an effective global charge}}.

\begin{figure*}
    \centering
    \includegraphics[width=0.95\textwidth]{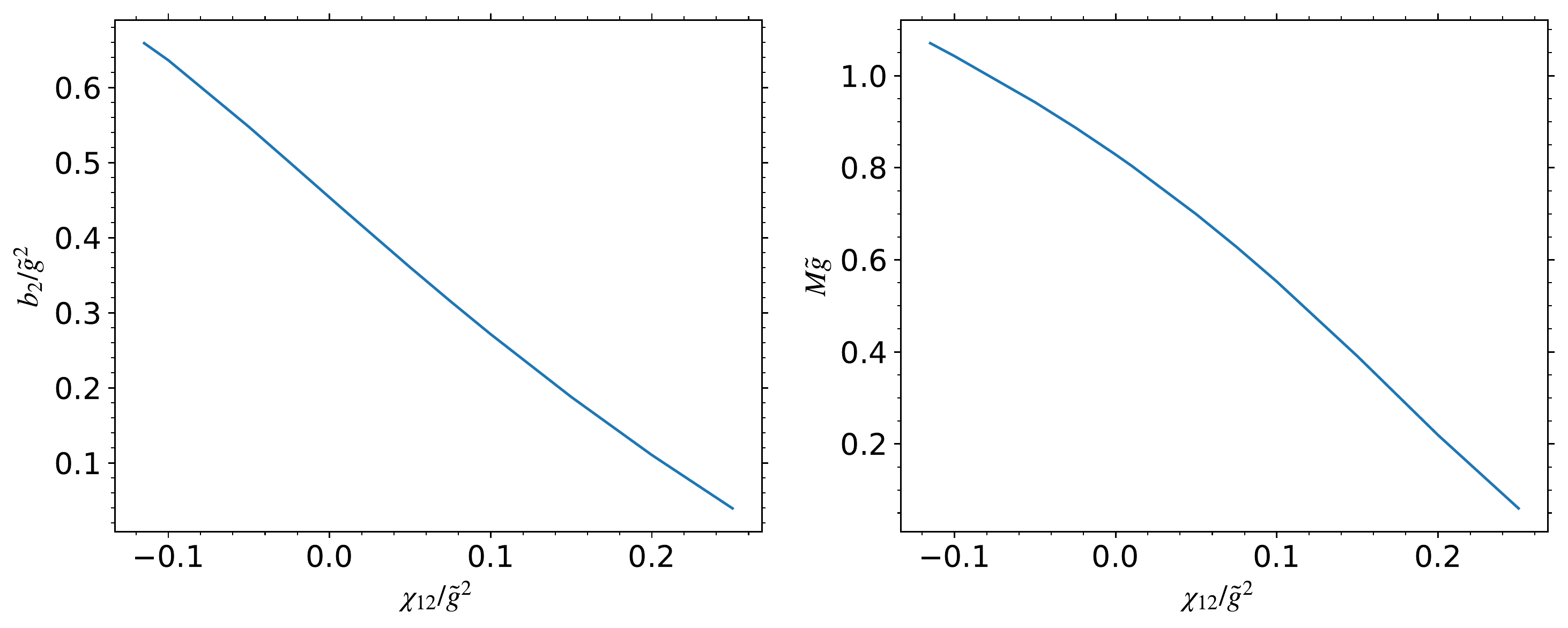}
    \caption{Equilibrium sequence of  the GSU2P particle-like \textcolor{black}{solution} with parameter $\chi_{12}$. \textcolor{black}{Left}: shooting parameter as a function of $\chi_6$. \textcolor{black}{Right}: gravitational mass as a function of $\chi_6$. There are no turning points.}
    \label{fig:sequence_C12}
\end{figure*}

The equilibrium sequence for one-node solutions is depicted in Fig. \ref{fig:sequence_C12}. Unfortunately, there are no turning points and, because the EYM soliton solution \textcolor{black}{has the feature of being unstable}, we conjecture that these new solutions may be unstable as well (see Sec. \ref{sec:stability}).
\subsubsection{Cases $\chi_5, \chi_6$}

\begin{figure*}
    \centering
    \includegraphics[width=0.95\textwidth]{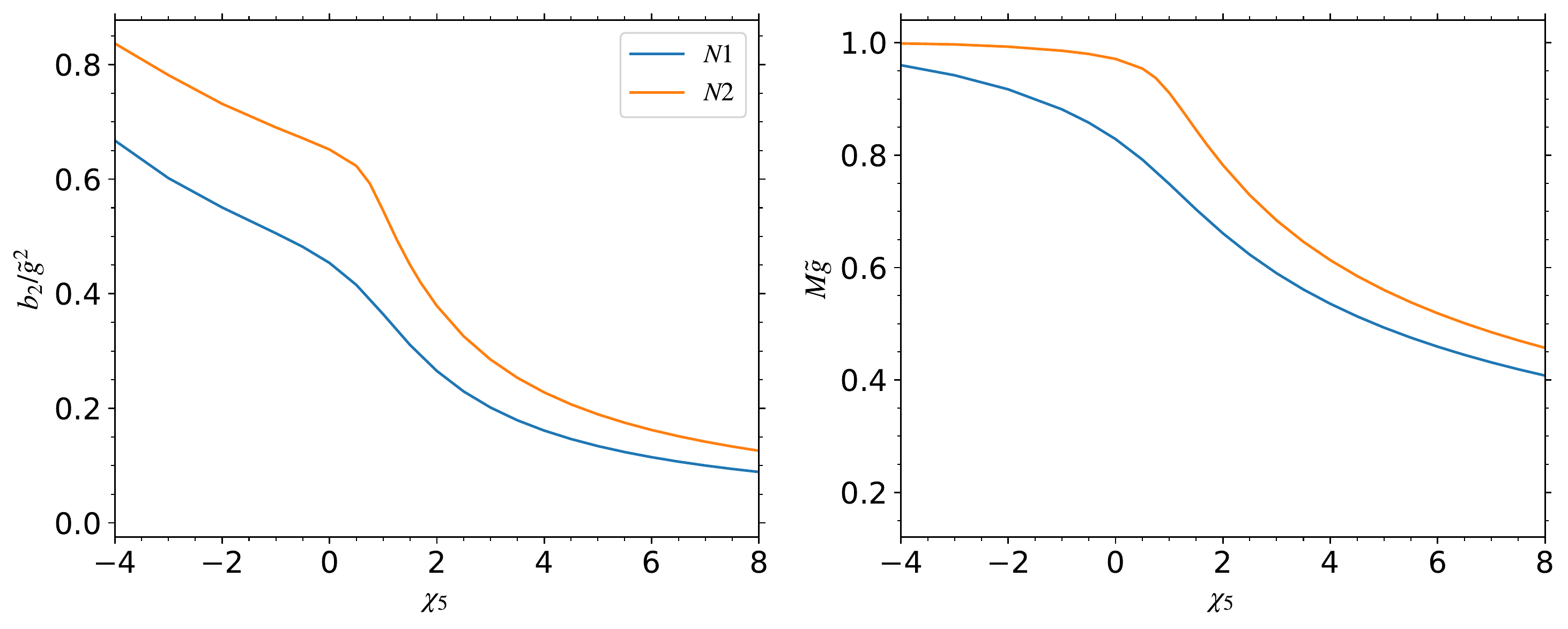}
    \caption{Equilibrium sequence  of the GSU2P particle-like \textcolor{black}{solutions} with \textcolor{black}{respect to the} parameter $\chi_{5}$ \textcolor{black}{for one-node (blue curve) and two-node families (orange curve).} \textcolor{black}{Left: shooting parameter as a function of $\chi_{5}$. Right: gravitational mass as a function of $\chi_{5}$}.}
    \label{fig:sequence_C5}
\end{figure*}

\begin{figure*}
    \centering
    \includegraphics[width=0.9\textwidth]{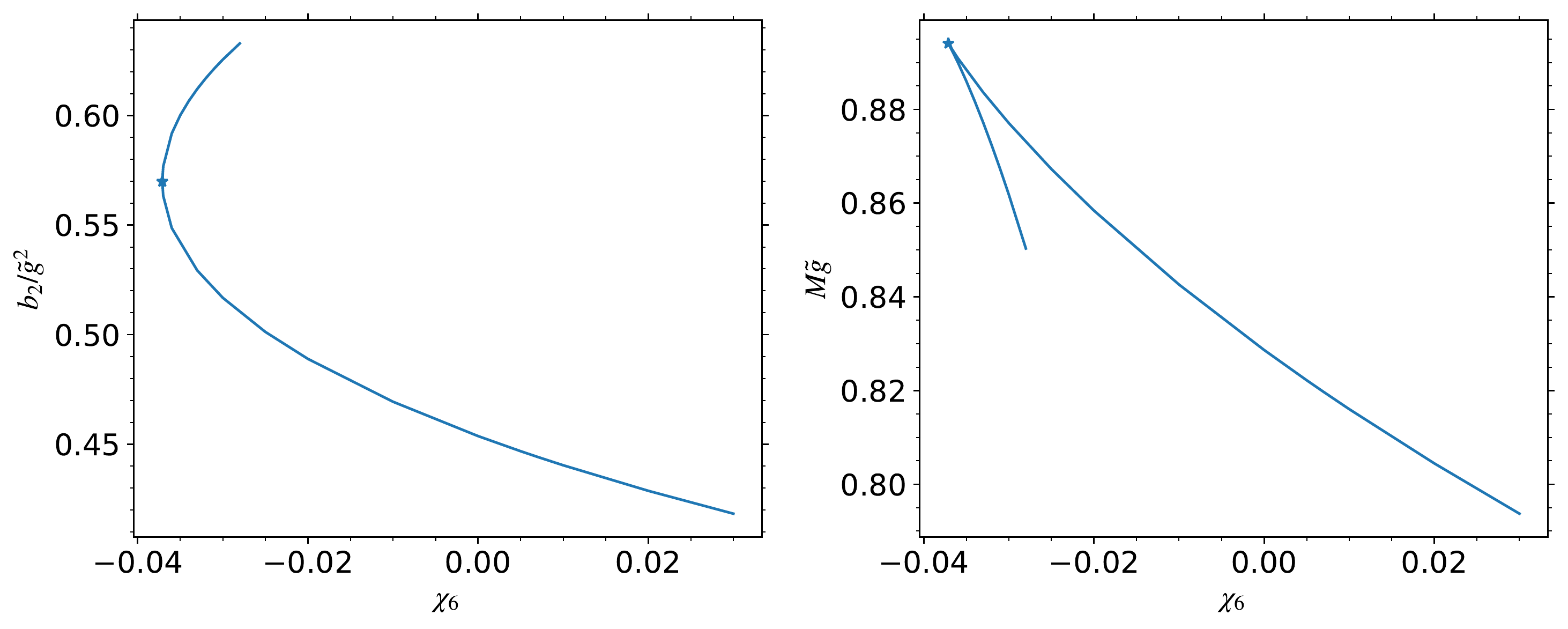}
    \caption{Equilibrium sequence of  GSU2P particle-like solutions  varying only $\chi_6$.  \textcolor{black}{Left}: shooting parameter as a function of $\chi_6$. \textcolor{black}{Right}: gravitational mass as a function of $\chi_6$. There is a turning point indicated by a star in both plots.}
\label{fig:sequence_C6}
\end{figure*}

The cases $\chi_5$ and $\chi_6$ include self-interactions involving derivatives of the vector field in the form $B^2 (\nabla B)^2$. However, in contrast to the previous case, this self-interaction does not give rise to globally charged solutions. 
The equilibrium sequences are shown in Figs. \ref{fig:sequence_C5} and \ref{fig:sequence_C6}. 

\begin{figure}
    \centering\includegraphics[width=0.49\textwidth]{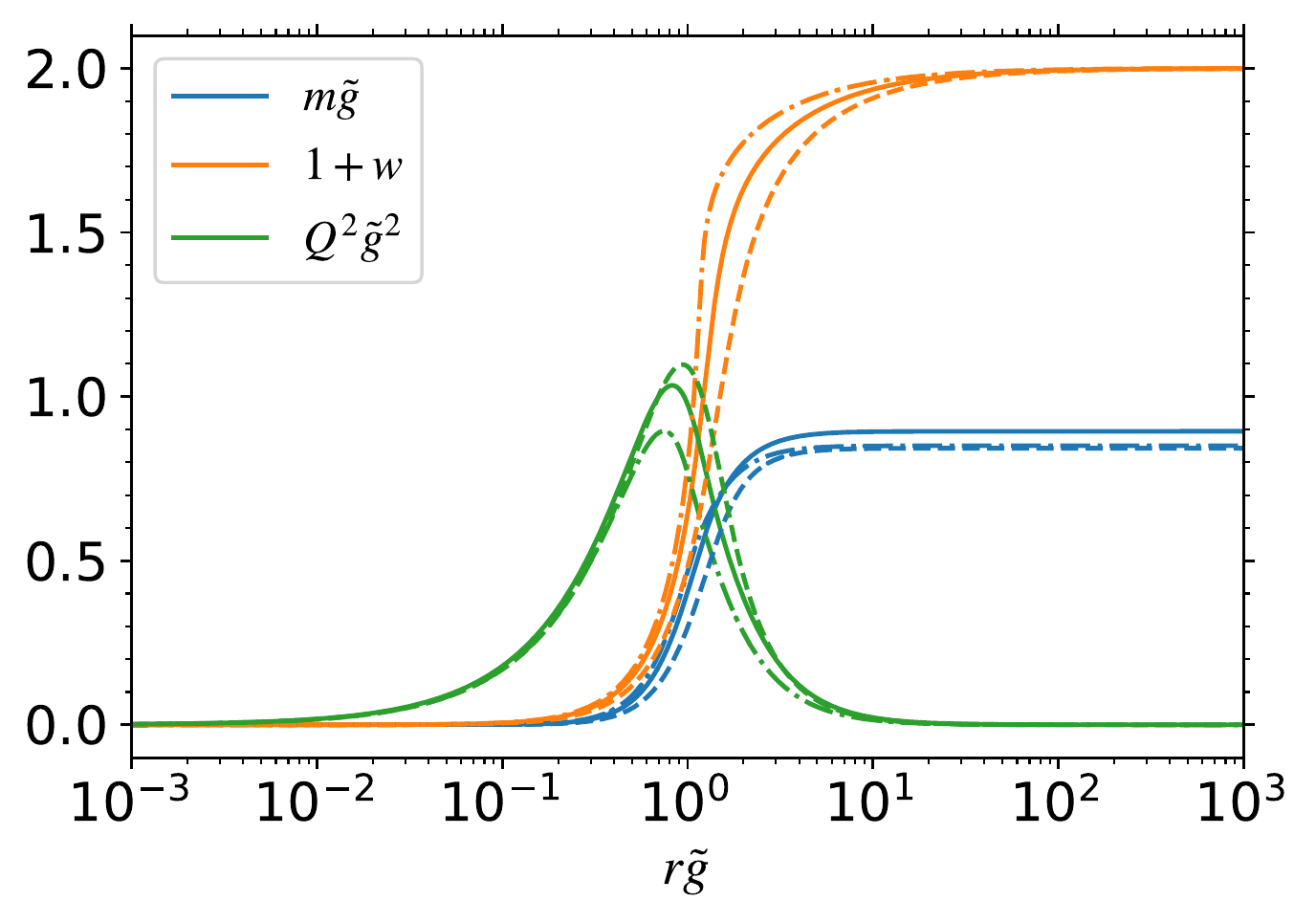}
    \caption{Mass function (blue), vector field (orange) and effective charge (green) of three representative members of one-node solutions of the  GSU2P $\chi_{6}$ family. The continuous curves correspond to the case $\chi_{6} = 0.0371$, i.e., the bifurcation point. The dotted-dashed curves correspond to the case $\chi_6= -0.028$, and the dashed curves to $\chi_6=-0.01$. Observe that in all cases the effective charge is real.}
\label{fig:PL_C6N1}
\end{figure}

The $\chi_5$ one-node and two-node families do not possess turning points, and the central density and gravitational mass grow as a function of $\chi_5$ \textcolor{black}{as can be seen in Fig. \ref{fig:sequence_C5}}. Furthermore, the energy density is positive, and the charge is real (see Fig. \ref{fig:PL_C6N1}). 

\begin{figure}
    \centering
    \includegraphics[width=0.47\textwidth]{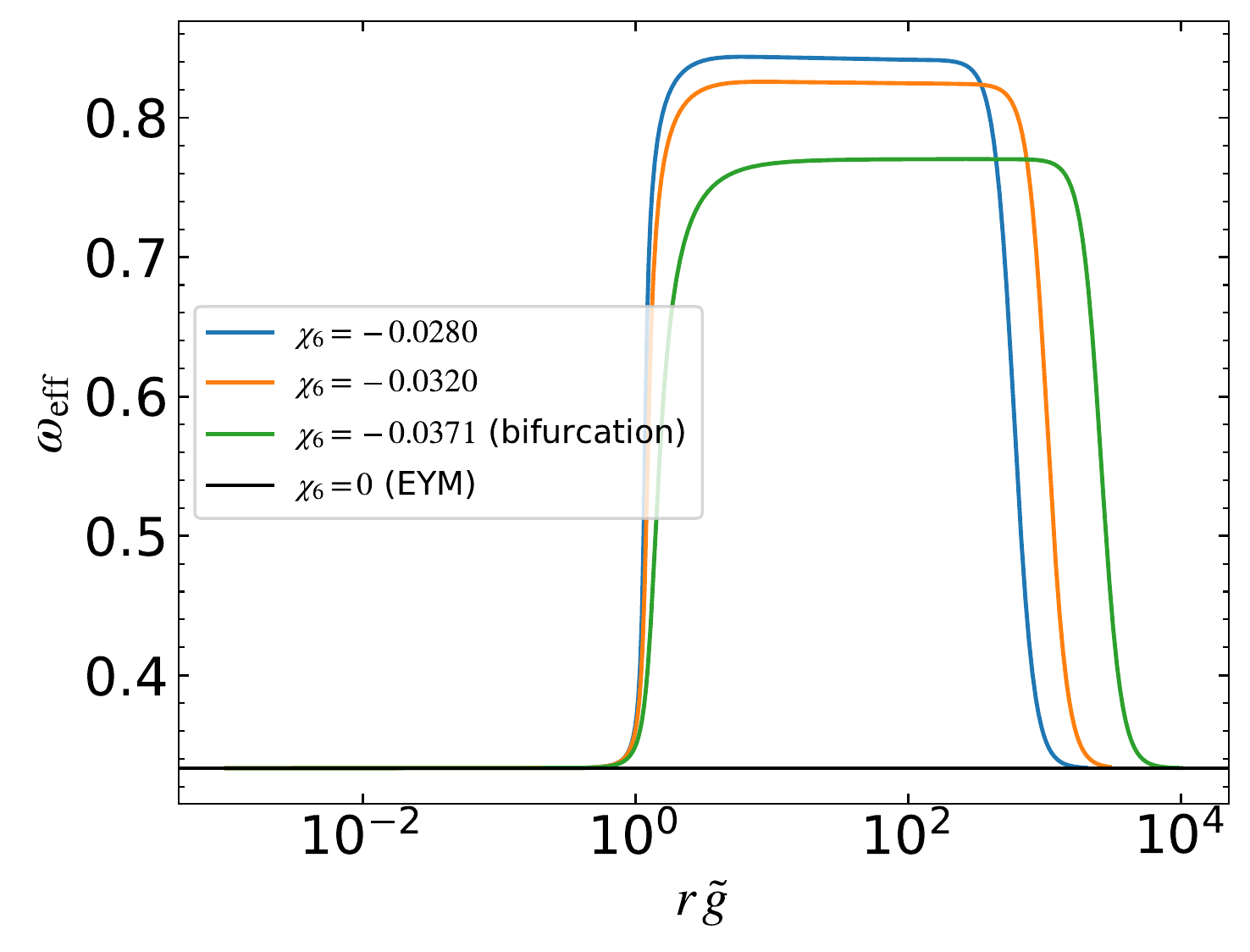}
    \caption{\textcolor{black}{Effective parameter of state $\omega_{\rm eff}\equiv p_{\rm eff}/\rho_{\rm eff}$ for some of the \emph{potentially stable} solutions of the $\chi_6$  one-node family. The state parameter is always positive but it behaves different from radiation in the transition zone.}}
    \label{fig:omega_C6}
\end{figure}

There is a turning point in the $\chi_6$ one-family located at $\chi_6\approx -0.0371$ \textcolor{black}{as can be appreciated in Fig. \ref{fig:sequence_C6}}. Unlike the non-minimal coupling case, the energy density is positive and the effective charge is real (for the latter, see Fig. \ref{fig:PL_C6N1}). \textcolor{black}{We also show, in  Fig. \ref{fig:omega_C6}, the state parameter $\omega_{\rm eff}$ of potentially stable one-node solutions. It can be seen that the fluid behaviour departs from the radiation-like one in the zone where the vector field is non trivial ($w \neq \pm 1$). Unlike the $\alpha_1$ case, the state parameter is always positive.} {\it \textcolor{black}{Thus, it seems a property of non-minimal couplings and quartic self-interactions to create regions with imaginary charges and negative energies}}.

\subsection{\textcolor{black}{Remarks on} stability}

\label{sec:stability}
The equilibrium and stability of stars in GR can be studied using a variational principle, see e.g. Refs. \cite{1965gtgc.book.....H,1988ApJ...325..722F,2000AstL...26..772S}. This procedure is related to the proper identification of the thermodynamical variables of the system, such as the total internal energy (gravitational mass), baryon number, entropy, and angular momentum. \textcolor{black}{Indeed}, thermodynamical variables can be used to deduce the stability properties of the equilibrium configurations in the case of black holes \cite{1993PhRvD..47.2234K}. This method is elegant and simple but requires careful selection of the potentials, the state and the control variables to predict the stability correctly, specially when non-minimal couplings are present \cite{2003PhRvD..68b4028T}.

Changes in the stability of radial modes occur at \textcolor{black}{the} turning points\footnote{Some authors e.g., Bizon \& Chmaj in Ref. \cite{1992PhLB..297...55B}, call turning points bifurcations. In contrast, other authors make a distinction between turning points (places where two or more sequences merge) and bifurcations (points where two or more sequences cut each other) \cite{1993PhRvD..47.2234K}.} in the equilibrium sequences. Since the EYM solutions are unstable, the existence of turning points is a necessary but not a sufficient condition to stabilize the solutions.  

In the case of the Einstein Non-Abelian Proca and Einstein-Skyrme scenarios \cite{1992PhLB..297...55B, 1995PhRvD..51.1510T}, catastrophe theory can be used to infer the stability of the solutions. The variables used there are the gravitational mass, the free parameters of the theory, and the shooting parameter.

Motivated by the results mentioned above, we have selected the gravitational mass, the free parameters of the theory, and the shooting parameter as the variables to speculate about the stability of the solutions. From this hypothesis, we conjecture that only the cases $\alpha_1$, $\alpha_3$, and $\chi_6$ are potential candidates for finding stable solutions. But, of course, the validity of our hypothesis can only be determined after performing a detailed analysis of the thermodynamical variables and the radial perturbations of the particle-like solutions.  We expect to address this analysis in a future work.

\section{Discussion} \label{sec:Discussion}

Do there exist gauge boson stars? If so, can they explain at least part of the dark matter content of the universe? Although there does not exist strong enough observational evidence yet regarding, mainly, the former question, one can gain some insight from the analysis of the predictions of well established frameworks such as GR or the Yang-Mills theories.  With no normal matter present, GR, which is the most successful gravitational theory we have to date, prohibits the existence of stationary, localized, and globally regular solutions.  This same conclusion is shared by the Yang-Mills theories which represents our most successful scheme up to now to describe the fundamental nuclear interactions.  In view of this, what does the EYM scenario have to say? The answer to this question was given by the seminal work of Bartnik and McKinnon \cite{1988PhRvL..61..141B} who demonstrated the existence of a family of particle-like solutions that, unfortunately, later on turned out to be unstable \cite{1990PhLB..237..353S}.  However, and having in mind that the desired stability might be related to additional contributions to the effective rest mass of the gauge fields, a new question is raised:  could a generalization of the Proca theory predict stable ``gauge'' boson stars? 

The main aim of this work has been to study the existence and properties of particle-like solutions in the framework of a modified theory of gravity that involves new gravitational degrees of freedom in the form of non-Abelian vector fields.  Such a theory is a generalization of the Proca theory when the action is invariant under global SU(2) internal transformations.  This theory, the GSU2P \cite{GallegoCadavid:2020dho,GallegoCadavid:2022uzn,Allys:2016kbq}, is a cousin of the well known\footnote{In the modified gravity circles.} generalized Proca theory developed a few years ago \cite{Tasinato:2014eka,Heisenberg:2014rta,Allys:2015sht,Allys:2016jaq,BeltranJimenez:2016rff,GallegoCadavid:2019zke} and which, in turn, is a cousin of the, well known as well, Horndeski's or Galileon theory \cite{1974IJTP...10..363H,deffayet1,Deffayet:2009mn,contrater}.  

Ignoring the $\mu$ and $\chi_4$ parameters in Eq. (\ref{eqn:lagrangian}), the GSU2P free of ghosts and Laplacian instabilities in the tensor sector, with gravity waves propagating at the speed of light in a cosmological background, is described by seven free coupling parameters: $\tilde{g}$, $\chi_1$, $\chi_2$, $\chi_5$, $\chi_6$, $\alpha_1$ and $\alpha_3$. 

Under the stationary pure magnetic configuration, also called the t'Hooft-Polyakov monopole, we have studied, analitically and numerically, the predictions of the GSU2P regarding the existence and properties of ``gauge'' boson stars. 

Analytical solutions have been found for the particular case where all the coupling parameters vanish except for $\tilde{g},\chi_1$, and $\chi_2$ together with a constant norm of the vector field. The two branch solutions correspond to the Reissner-Nordstr\"om spacetime with different effective charges which can be real or imaginary depending on the specific values of $\chi_1$ and $\chi_2$.

The full scenario with all the seven coupling constants different to zero could only be analyzed numerically.  To this end, we built asymptotic series both at the origin and at infinity searching for conditions that guaranteed the regularity and asymptotic flatness of the solutions.  Then, we solved the field equations following the shooting parameter technique.  The shooting parameter turned out to be related to the central energy density of the star.  As a result, both globally neutral and globally charged solutions emerged in the form of one-parameter equilibrium sequences.  Some of these solutions exhibit positive energy density and local real effective charge while others exhibit negative energy density and local imaginary effective charge.  These are new features of the GSU2P compared to the EYM case.  Another new feature is the existence of points where two sequences of one-node solutions merge, called turning points.

\textcolor{black}{In addition to the merging of different families of particle-like solutions, ongoing work on black hole solutions in the GSU2P \cite{inprogress} shows that when the areal-radius of the horizon goes to 0, the black hole family merges with a soliton family. Also, we have extended and analyzed the analytical black hole solution of section \ref{sec:analytical} in Ref. \cite{Gomez:2023wei}.}

Our work has allowed us to reach two important and interesting conclusions:  the creation of an effective global charge is the result of the new non-derivative self-interactions that are present in the GSU2P and the regions with negative energy density and local imaginary charge seem to be the result of the presence of non-minimal couplings and quartic self-interactions which are also  characteristics of the GSU2P. 

We cannot say anything definitive about the stability of the found solutions yet.  However, since the existence of turning points in the equilibrium sequences is a necessary but not sufficient condition for stability, we can claim that some of the found solutions might indeed be stable and become the first examples of theoretically viable ``gauge'' boson stars. For instance, in the case of the $\alpha_1$ one-node family, potential stable solutions constraint $\alpha_1$ to be in the interval $(0.0137,0.0295)$. Of course, this is a subject we expect to concentrate on in the near future.

\section*{Acknowledgements} \label{sec:acknowledgements}

 J. N. M., J. F. R., and Y. R. have received funding/support from the Patrimonio Autónomo - Fondo Nacional de Financiamiento para la Ciencia, la Tecnología y la Innovación Francisco José de Caldas (MINCIENCIAS—COLOMBIA) Grant No. 110685269447 RC-80740-465-2020, Project No. 69553.  Y. R. has received funding/support from the European Union’s Horizon 2020 research and innovation programme under the Marie Skłodowska-Curie Grant No.860881-HIDDeN. G. G. acknowledges financial support from Agencia Nacional de Investigaci\'on y Desarrollo (ANID) through the FONDECYT postdoctoral Grant No. 3210417.


\bibliography{References/referencias,References/references_post,References/references,References/references2,References/referencesPhD,References/sample,References/gcn}

\appendix
\section{Covariant field equations} \label{sec:appendixFE}
The field equations are highly complicated.  Here, we present their explicit form by giving the contribution of each term in the action. The equations obtained by varying the action with respect to $g^{\mu\nu}$ have the following form,
\begin{equation}
    G_{\mu\nu} +\mathcal{G}_{\mu\nu}^{\rm canonical}+ G_{\mu\nu}^{\rm GSU2P} = G_{\mu\nu} -8\pi T_{\mu\nu}^{\rm eff}=0\, ,
\end{equation}
where
\begin{multline}
\mathcal{G}_{\mu\nu}^{\rm canonical} \equiv \tfrac{1}{2} F^2 g_{\mu \nu }-2 F_{a\nu \alpha } F^{a}{}_{\mu }{}^{\alpha } -2 \mu^2 X g_{\mu \nu } \\
+ 2 \mu^2 f_{\mu \nu } + 4 \tilde{g} \epsilon_{abc} B^{a\alpha } B^{b}{}_{(\mu } F^{c}{}_{\nu) \alpha } \\
+ 4 B_{a}{}_{(\mu } \nabla^{\alpha}F^{a}{}_{\nu) }{}_{\alpha} \,,
\end{multline}
is the canonical non-Abelian Proca contribution, and the contributions from the generalized Proca SU(2) terms are given by
\begin{equation}
    G_{\mu\nu}^{\rm GSU2P} \equiv \sum_{n=1}^6\alpha_n \mathcal{G}_{\mu\nu}^n+\sum_{n=1}^7\chi_n \mathcal{H}_{\mu\nu}^n \,,
\end{equation}
\begin{equation}
    \mathcal{G}^n_{\mu\nu} \equiv \frac{1}{\sqrt{-g}} \frac{\delta(\sqrt{-g}\mathcal{L}^n_{4,2})}{\delta g^{\mu\nu}}\, ,
\end{equation}
\begin{equation}
    \mathcal{H}^n_{\mu\nu} \equiv \frac{1}{\sqrt{-g}} \frac{\delta(\sqrt{-g}\mathcal{L}^n_{2})}{\delta g^{\mu\nu}}\, ,
\end{equation}
\begin{multline}
    \mathcal{G}^1_{\mu\nu} \equiv f_{\mu \nu } ( S^2- S_{a}{}^{\beta }{}_{\beta } S^{a\alpha }{}_{\alpha }) + S^{a}{}_{\mu \nu } (2 S_{a}{}^{\alpha }{}_{\alpha } X\\
    - 2 B^{a\alpha } B^{b}{}_{\alpha } S_{b}{}^{\beta }{}_{\beta }-4 B_{a}{}^{\alpha } B^{b\beta } S_{b\alpha \beta } -2 f^{\alpha \beta } S_{a\alpha \beta }) \\
     + B^{a}{}_{\mu } B^{b}{}_{\nu } (2 S_{a}{}^{\alpha \beta } S_{b\alpha \beta } - 2 S_{a}{}^{\alpha }{}_{\alpha } S_{b}{}^{\beta }{}_{\beta }) \\
    - 4 B^{a\alpha } B_{b}{}_{\alpha } A_{a(\mu }{}^{\beta } S^{b}{}_{\nu) \beta } + 4 X A_{a}{}_{(\mu }{}^{\alpha } S^{a}{}_{\nu) \alpha } \\
    + g_{\mu \nu } \bigl[2 f^{\alpha \beta } S_{a}{}^{\gamma }{}_{\gamma } S^{a}{}_{\alpha \beta } + B^{a\alpha } B^{b}{}_{\alpha } S_{a}{}^{\beta \gamma } S_{b\beta \gamma } \\
    + 4 B^{a\alpha } B^{b\beta } S_{a\alpha \beta } S_{b}{}^{\gamma }{}_{\gamma } + B^{a\alpha } B^{b}{}_{\alpha } S_{a}{}^{\beta }{}_{\beta } S_{b}{}^{\gamma }{}_{\gamma } \\
    - X (S_{a}{}^{\beta }{}_{\beta } S^{a\alpha }{}_{\alpha } + S_{a\alpha \beta } S^{a\alpha \beta } + 4 B^{a\alpha } \nabla_{\alpha }S_{a}{}^{\beta }{}_{\beta }) \\
    + 4 B^{a\alpha } f_{\alpha }{}^{\beta } \nabla_{\beta }S_{a}{}^{\gamma }{}_{\gamma }\bigr]+ 4 B^{a\alpha } \nabla_{\alpha }S_{a\mu \nu } \\
    - 4 B^{a\alpha } f_{\alpha }{}^{\beta } \nabla_{\beta }S_{a\mu \nu } \,,
\end{multline}
\begin{multline}
\mathcal{G}^2_{\mu\nu} \equiv g_{\mu \nu }(B^{a\alpha } B^{b\beta } A_{a\alpha }{}^{\gamma }  S_{b\beta \gamma }+ \tfrac{1}{2} B^{a\alpha } B^{b\beta } A_{a\alpha }{}^{\gamma } A_{b\beta \gamma }  \\
-  \tfrac{1}{2} B^{a\alpha } B^{b\beta } A_{a\beta }{}^{\gamma }  S_{b\alpha \gamma } -  \tfrac{1}{2} A_{a\alpha }{}^{\gamma } A^{a\alpha \beta } f_{\beta \gamma } -  \tfrac{1}{2} A^{a\alpha \beta } f_{\alpha }{}^{\gamma }  S_{a\beta \gamma }) \\
+ \tfrac{1}{2} f_{\mu \nu }( S^2 -   S_{a}{}^{\beta }{}_{\beta } S^{a\alpha }{}_{\alpha }) - \tfrac{1}{2} A_{a\alpha }{}^{\beta } A^{a}{}_{(\mu }{}^{\alpha } f_{\nu) \beta } \\
+ \tfrac{1}{2} B^{a}{}_{\mu } B^{b}{}_{\nu }( S_{a}{}^{\alpha }{}_{\alpha } S_{b}{}^{\beta }{}_{\beta } -   S_{a}{}^{\alpha \beta } S_{b\alpha \beta }) -B^{b\alpha }B_{a}{}_{(\mu } A^{a}{}_{\nu) }{}^{\beta }  A_{b\alpha \beta } \\
-  \tfrac{1}{2} B^{a\alpha } B^{b\beta } A_{a\mu \beta } A_{b\nu \alpha } + \tfrac{1}{2}  B_{b}{}^{\alpha } A_{a\alpha \beta }B^{a}{}_{(\mu } A^{b}{}_{\nu) }{}^{\beta } + \tfrac{1}{2} A_{a\nu }{}^{\beta } A^{a}{}_{\mu }{}^{\alpha } f_{\alpha \beta } \\
+ A^{a}{}_{(\mu }{}^{\alpha } f_{\nu) }{}^{\beta } S_{a\alpha \beta } + \tfrac{1}{2}  B_{b}{}^{\alpha }B^{a}{}_{(\mu } A^{b}{}_{\nu) \alpha } S_{a}{}^{\beta }{}_{\beta } -\tfrac{1}{2} A^{a}{}_{(\mu }{}^{\alpha } f_{\nu) \alpha } S_{a}{}^{\beta }{}_{\beta } \\
+ \tfrac{1}{2}  B^{b\alpha } A_{b\alpha }{}^{\beta }B_{a}{}_{(\mu } S^{a}{}_{\nu) \beta } -  \tfrac{1}{2} A_{a\alpha}{}^{\beta } f_{(\mu}{}^{ \alpha } S^{a}{}_{\nu)\beta } - \tfrac{1}{2}  f_{\alpha }{}^{\beta }A_{a}{}_{(\mu }{}^{\alpha } S^{a}{}_{\nu) \beta }\\
+  S_{a}{}^{\beta }{}_{\beta }f_{(\mu }{}^{\alpha }  S^{a}{}_{\nu) \alpha } -   S_{a\alpha \beta }f_{(\mu }{}^{\alpha } S^{a}{}_{\nu )}{}^{\beta } -  B^{b\alpha } B_{a}{}_{(\mu }A^{a}{}_{\nu)}{}^{\beta } S_{b\alpha \beta } \\
+ \tfrac{1}{2} B^{b\alpha } B_{a}{}_{(\mu } S^{a}{}_{\nu) }{}^{\beta } S_{b\alpha \beta }  - \tfrac{1}{2}  B^{b\alpha }B_{a}{}_{(\mu } S^{a}{}_{\nu) \alpha } S_{b}{}^{\beta }{}_{\beta }\\
-  \tfrac{1}{2} B_{b}{}^{\alpha }  B^{a}{}_{(\mu } S^{b}{}_{\nu) \alpha }S_{a}{}^{\beta }{}_{\beta } -  \tfrac{1}{2} B_{a}{}^{\alpha } B_{b\beta } A^{a}{}_{(\mu}{}^{\beta } S^{b}{}_{\nu) \alpha }  \\
+ B^{a}{}_{\alpha } B_{b}{}^{\beta } A_{a(\mu}{}^{ \alpha } S^{b}{}_{\nu) \beta } + \tfrac{1}{2} B_{b}{}^{\alpha } S_{a\alpha \beta } B^{a}{}_{(\mu } S^{b}{}_{\nu) }{}^{\beta } \\
- f_{\mu \nu } \bigr( B^{a\alpha } \nabla_{\alpha}S_{a}{}^{\beta }{}_{\beta } - B^{a\alpha } f_{\mu \nu } \nabla_{\beta}S_{a\alpha }{}^{\beta }\bigl) +  B^{a}{}_{(\mu } f_{\nu) }{}^{\alpha } \nabla_{\alpha}S_{a}{}^{\beta }{}_{\beta } \\
+  B_{a}{}_{\alpha } f^{\beta }{}_{(\mu } \nabla^{\alpha}S^{a}{}_{\nu) \beta }-   f^{\alpha}{}_{\beta } B_{a}{}_{(\mu }\nabla^{\beta }A^{a}{}_{\nu) \alpha } + B_{a}{}^{\alpha } f_{\beta }{}_{(\mu } \nabla^{\beta}A^{a}{}_{\nu) \alpha } \\
-  B^{a}{}_{(\mu } f_{\nu) }{}^{\alpha} \nabla_{\beta}S_{a\alpha }{}^{\beta } - B_{a}{}^{\alpha } f_{\beta }{}_{(\mu } \nabla^{\beta}S^{a}{}_{\nu) \alpha } \\
+ B^{a\alpha } f^{\beta \gamma } g_{\mu \nu } \nabla_{\gamma }A_{a\alpha \beta } \,,
\end{multline}
\begin{multline}
\mathcal{G}^3_{\mu\nu} \equiv 3 X^2 G_{\mu \nu }  + 3 X f_{\mu \nu } R + \tfrac{1}{2} g_{\mu \nu }f^{\alpha \beta } f^{\gamma \delta }  R_{\alpha \gamma \beta \delta } \\
+ f^{\beta \gamma } f_{(\mu }{}^{\alpha } R_{\nu) \beta \alpha \gamma } 
-\nabla_{\alpha}\nabla_{\beta}(f^{\alpha\beta}f_{\mu\nu})\\
+\tfrac{1}{2}\nabla_{\alpha}\nabla_{\beta}(f_{\mu}{}^{\alpha}f^{\beta}{}_{\nu}) +\tfrac{1}{2}\nabla_{\beta}\nabla_{\alpha}(f_{\mu}{}^{\alpha}f^{\beta}{}_{\nu})\\
+  3 g_{\mu \nu } \Box X^2 - 3 \nabla_{\mu}\nabla_{\nu}X^2 \,,     
\end{multline}
\begin{multline}
    \mathcal{G}^4_{\mu\nu} \equiv \bigl(4 X^2  + 2  f_{\alpha \beta } f^{\alpha \beta }\bigr)G_{\mu \nu } + 4\bigl( X f_{\mu \nu } - f_{\mu }{}^{\alpha } f_{\nu \alpha } \bigr) R\\
    + 2g_{\mu\nu}\bigl[ 2 \Box X^2+ \Box (f_{\alpha \beta }f^{\alpha\beta })\bigr] \\
    - 4 \nabla_{\mu }\nabla_{\nu}X^2 - 2   \nabla_{\mu }\nabla_{\nu} (f_{\alpha \beta }f^{\alpha \beta }) \,,
\end{multline}
\begin{multline}
    \mathcal{G}^5_{\mu\nu} \equiv  - 2 X^2 G_{\mu \nu }  - f_{\mu \nu } R   -(f_{\mu\nu} - g_{\mu \nu } X) G^{\alpha \beta } f_{\alpha \beta } \\
    -g_{\mu\nu}[X^2 R + \nabla_{\alpha}\nabla_{\beta }(Xf^{\alpha \beta } ) - 2 \Box X^2 ] \\
    + 2 \nabla_{\alpha}\nabla_{(\mu}[ X f_{\mu)}{}^{\alpha}]- \Box(X f_{\mu\nu}) +  2\nabla_{\mu}\nabla_{\nu}X^2 \,,
\end{multline}
\begin{multline}
    \mathcal{G}^6_{\mu\nu} \equiv - \tfrac{1}{2}  f_{\alpha \beta }  f^{\alpha \beta }G_{\mu \nu }  -  (f_{\mu \alpha } f_{\nu \beta } + \tfrac{1}{2}g_{\mu\nu} f_{\alpha }{}^{\gamma } f_{\beta \gamma } )G^{\alpha \beta }\\
     - \tfrac{1}{2} g_{\mu\nu}\bigl[\tfrac{1}{2}  f_{\alpha \beta }  f^{\alpha \beta } R -  \nabla_{\alpha}\nabla_{\beta}(f^{\alpha\gamma}f^{\beta}{}_{\gamma}) +\Box(f_{\alpha\beta}f^{\alpha\beta})\bigr]\\
    + \tfrac{1}{2} f_{\mu }{}^{\alpha } f_{\nu \alpha } R -\tfrac{1}{2}\nabla_{\alpha}\nabla_{(\mu}[f_{\nu)}{}^{\gamma}f^{\alpha}{}_{\gamma}]\\
    +\tfrac{1}{2}\Box(f_{\mu}{}^{\gamma}f_{\nu\gamma}) +\tfrac{1}{2}\nabla_{\mu}\nabla_{\nu}(f_{\alpha\beta}f^{\alpha\beta}) \,,
\end{multline}
\begin{equation}
    \mathcal{H}^1_{\mu\nu} \equiv 4 X f_{\mu \nu } -2 X^2 g_{\mu \nu } \,, 
\end{equation}
\begin{equation}
    \mathcal{H}^2_{\mu\nu} \equiv -2 f_{\mu }{}^{\alpha } f_{\nu \alpha } -  \tfrac{1}{2} f_{\alpha \beta } f^{\alpha \beta } g_{\mu \nu } \,, 
\end{equation}
\begin{multline}
    \mathcal{H}^3_{\mu\nu} \equiv  A_{a}{}^{\alpha \beta } A_{b\alpha \beta }B^{a}{}_{\mu } B^{b}{}_{\nu } +  f_{\alpha \beta } A_{a\nu }{}^{\beta } A^{a}{}_{\mu }{}^{\alpha }\\
    -  \tfrac{1}{2} A_{a\alpha }{}^{\gamma } A^{a\alpha \beta } f_{\beta \gamma } g_{\mu \nu } + B_{b}{}^{\alpha } A_{a\alpha \beta } B^{a}{}_{(\mu } A^{b}{}_{\nu) }{}^{\beta }\\
    -  B^{b\alpha }  A_{b\alpha \beta }B_{a}{}_{(\mu } A^{a}{}_{\nu) }{}^{\beta } - B^{b\alpha } S_{b\alpha \beta }B_{a}{}_{(\mu } A^{a}{}_{\nu) }{}^{\beta }  \\
    -   B^{b\alpha }  S_{b}{}^{\beta }{}_{\beta } B_{a}{}_{(\mu }A^{a}{}_{\nu) \alpha } -  B_{b}{}^{\alpha } A_{a\alpha }{}^{\beta }B^{a}{}_{(\mu } S^{b}{}_{\nu) \beta } \\
    -2 B^{a}{}_{(\mu } f_{\nu )}{}^{\alpha } \nabla_{\beta }A_{a\alpha }{}^{\beta } - 2  f^{\alpha}{}_{\beta }B_{a}{}_{(\mu } \nabla^{\beta }A^{a}{}_{\nu )\alpha } \,,
\end{multline}
\begin{multline}
    \mathcal{H}^4_{\mu\nu} \equiv B^{a}{}_{\mu } B^{b}{}_{\nu } A_{a}{}^{\alpha \beta } A_{b\alpha \beta }+ B^{a\alpha } B^{b\beta } A_{a\mu \beta } A_{b\nu \alpha }\\
    - B_{b}{}^{\alpha } A_{a\alpha \beta }B^{a}{}_{(\mu } A^{b}{}_{\nu) }{}^{\beta } -   B_{b}{}^{\alpha } S_{a\alpha \beta } B^{a}{}_{(\mu }A^{b}{}_{\nu) }{}^{\beta }\\
    -   A_{a\alpha }{}^{\beta } A^{a}{}_{(\mu }{}^{\alpha } f_{\nu) \beta } -   S_{a}{}^{\beta }{}_{\beta } A^{a}{}_{(\mu }{}^{\alpha } f_{\nu) \alpha }\\
    -  \tfrac{1}{2} B^{a\alpha } B^{b\beta } A_{a\beta }{}^{\gamma } A_{b\alpha \gamma } g_{\mu \nu } -  A_{a}{}^{\alpha \beta } f_{\alpha (\mu} S^{a}{}_{\nu )\beta } \\
    - 2 B_{a}{}_{\alpha } f^{\beta }{}_{(\mu } \nabla^{\alpha }A^{a}{}_{\nu) \beta } -  2 B^{a}{}_{(\mu } f_{\nu) }{}^{\alpha } \nabla_{\beta }A_{a\alpha }{}^{\beta } \,,
\end{multline}
\begin{multline}
    \mathcal{H}^5_{\mu\nu} \equiv B^{a\alpha } B^{b\beta } A_{a\mu \alpha } A_{b\nu \beta } -  \tfrac{1}{2} B^{a\alpha } B^{b\beta } A_{a\alpha }{}^{\gamma } A_{b\beta \gamma } g_{\mu \nu } \\
    + f_{\mu \nu } (A_{a\alpha \beta } A^{a\alpha \beta } - 2 B^{a\alpha } \nabla_{\beta }A_{a\alpha }{}^{\beta }) \\
    +  A_{a\alpha }{}^{\beta } A^{a}{}_{(\mu }{}^{\alpha } f_{\nu) \beta } + B^{b\alpha } A_{b\alpha \beta} B_{a}{}_{(\mu } A^{a}{}_{\nu)}{}^{\beta }  \\
    -  A^{a}{}_{(\mu }{}^{\alpha } f_{\nu) }{}^{\beta } S_{a\alpha \beta } - B_{b}{}^{\alpha } S_{a}{}^{\beta }{}_{\beta } B^{a}{}_{(\mu } A^{b}{}_{\nu) \alpha }\\
    -  B^{b\alpha } A_{b\alpha }{}^{\beta } B_{a}{}_{(\mu } S^{a}{}_{\nu) \beta } - 2 B_{a}{}^{\alpha } f_{\beta }{}_{(\mu }\nabla^{\beta }A^{a}{}_{\nu) \alpha } \,,
\end{multline}
\begin{multline}
    \mathcal{H}^6_{\mu\nu} \equiv -  A_{a\alpha \beta } A^{a\alpha \beta } f_{\mu \nu } + X A_{a\alpha \beta } A^{a\alpha \beta } g_{\mu \nu }\\+
    4 B^{b\alpha } A_{b\alpha \beta }B_{a}{}_{(\mu } A^{a}{}_{\nu) }{}^{\beta } - 4  B^{b\alpha } S_{b\alpha \beta } B_{a}{}_{(\mu } A^{a}{}_{\nu) }{}^{\beta }\\
    - 4 X  A^{a}{}_{\mu }{}^{\alpha } A_{a\nu \alpha } + 8 X B_{a}{}_{(\mu } \nabla_{\alpha }A^{a}{}_{\nu)}{}^{\alpha } \,,
\end{multline}
\begin{multline}
    \mathcal{H}^7_{\mu\nu} \equiv 2 B^{a\alpha } B^{b}{}_{\alpha } A_{a\mu }{}^{\beta } A_{b\nu \beta }- B^{a}{}_{\mu } B^{b}{}_{\nu } A_{a}{}^{\alpha \beta } A_{b\alpha \beta } \\
    + 2 B_{b}{}^{\alpha } A_{a\alpha \beta } B^{a}{}_{(\mu } A^{b}{}_{\nu) }{}^{\beta } -  2 A_{a\alpha }{}^{\beta } A^{a}{}_{(\mu }{}^{\alpha } f_{\nu) \beta } \\
    -  \tfrac{1}{2} B^{a\alpha } B^{b}{}_{\alpha } A_{a}{}^{\beta \gamma } A_{b\beta \gamma } g_{\mu \nu } -  2 B_{b}{}^{\alpha } S_{a\alpha \beta } B^{a}{}_{(\mu }A^{b}{}_{\nu) }{}^{\beta } \\
    - 2 S_{a\alpha \beta } A^{a}{}_{(\mu }{}^{\alpha } f_{\nu) }{}^{\beta } - 4 B_{a}{}^{\alpha } f_{\alpha (\mu } \nabla^{\beta }A^{a}{}_{\nu) }{}_{\beta } \,,
\end{multline}
where $X\equiv -B_{a\mu}B^{a\mu}/2$,  $f_{\mu\nu}\equiv B^a{}_{\mu} B_{a\nu}$,   $\square\equiv \nabla_{\mu}\nabla^{\mu}$,  $S^2\equiv S_{a\alpha\beta}S^{a\alpha\beta}$, $A^2\equiv A_{a\alpha\beta}A^{a\alpha\beta}$, \textcolor{black}{$F^2\equiv F_{a\alpha\beta}F^{a\alpha\beta}$}, and the symmetrization and antisymmetrization  operations are defined as  $Y_{(\mu\nu)}\equiv\left( Y_{\mu\nu}+Y_{\nu\mu}\right)/2$, $Y_{[\mu\nu]}\equiv\left( Y_{\mu\nu}-Y_{\nu\mu}\right)/2$.

The field equations obtained by varying the action with respect to $B^{a\mu}$ have the following form:
\begin{equation}
    K_{a\mu} \equiv K_{a\mu}^{\rm canonical}+\sum_{n=1}^6\alpha_n K_{a\nu}^n+\sum_{n=1}^7\chi_n J_{a\mu}^n =0\,,
\end{equation}
where
\begin{equation}
    K_{a\mu}^{\rm canonical} \equiv -\mu^2 B_{a\mu } - \tilde{g} B^{b\alpha } \epsilon_{abc} F^{c}{}_{\mu \alpha } - \nabla_{\alpha }F_{a\mu }{}^{\alpha } \,,
\end{equation}
\begin{equation}
    K^n_{a\mu} \equiv \frac{1}{4}\frac{\delta \mathcal{L}_{4,2}^n}{\delta B^{a\mu}} \,,
\end{equation}
\begin{equation}
    J^n_{a\mu} \equiv \frac{1}{4}\frac{\delta \mathcal{L}_{2}^n}{\delta B^{a\mu}} \,,
\end{equation}
\begin{multline}
    K_{a\mu}^1 \equiv B^{b}{}_{\mu } (S_{a}{}^{\alpha }{}_{\alpha } S_{b}{}^{\beta }{}_{\beta } - S_{a}{}^{\alpha \beta } S_{b\alpha \beta }) \\
    -  B^{b\alpha } (A_{b\mu \alpha } S_{a}{}^{\beta }{}_{\beta } + A_{b\alpha }{}^{\beta } S_{a\mu \beta } -  S_{a\mu }{}^{\beta } S_{b\alpha \beta }\\
    + A_{a\mu \alpha } S_{b}{}^{\beta }{}_{\beta } + S_{a\mu \alpha } S_{b}{}^{\beta }{}_{\beta } + S_{a}{}^{\beta }{}_{\beta } S_{b\mu \alpha }\\
    + A_{a\alpha }{}^{\beta } S_{b\mu \beta }-  S_{a\alpha }{}^{\beta } S_{b\mu \beta })\\
    + \tfrac{1}{2} B_{a\mu } (S_{b}{}^{\beta }{}_{\beta } S^{b\alpha }{}_{\alpha } -  S^2) + 2 X ( \nabla_{\mu }S_{a}{}^{\alpha }{}_{\alpha } - \nabla_{\alpha }S_{a\mu }{}^{\alpha }) \\
    -  B_{a}{}^{\alpha } (A^{b}{}_{\mu \alpha } S_{b}{}^{\beta }{}_{\beta } + A^{b}{}_{\alpha }{}^{\beta } S_{b\mu \beta } + S_{b}{}^{\beta }{}_{\beta } S^{b}{}_{\mu \alpha }\\
    -  S_{b\alpha \beta } S^{b}{}_{\mu }{}^{\beta } - 2 B^{b}{}_{\alpha } \nabla_{\beta }S_{b\mu }{}^{\beta } + 2 B^{b}{}_{\alpha } \nabla_{\mu }S_{b}{}^{\beta }{}_{\beta }) \,,
\end{multline}
\begin{multline}
    K_{a\mu}^2 \equiv \tfrac{1}{8} \bigl[2 B_{a}{}^{\alpha } A_{b\alpha \beta } A^{b}{}_{\mu }{}^{\beta } - B_{a\mu }  A^2 - 2 B_{a}{}^{\alpha } A^{b}{}_{\mu }{}^{\beta } S_{b\alpha \beta } \\
    + 2 B_{a}{}^{\alpha } A^{b}{}_{\mu \alpha } S_{b}{}^{\beta }{}_{\beta } + 2 B_{a}{}^{\alpha } A^{b}{}_{\alpha }{}^{\beta } S_{b\mu \beta } + B_{a\mu } S_{b}{}^{\beta }{}_{\beta } S^{b\alpha }{}_{\alpha }\\
    -  B_{a\mu } S_{b\alpha \beta } S^{b\alpha \beta } - 2 B_{a}{}^{\alpha } S_{b}{}^{\beta }{}_{\beta } S^{b}{}_{\mu \alpha } + 2 B_{a}{}^{\alpha } S_{b\alpha \beta } S^{b}{}_{\mu }{}^{\beta }\\
    - 2 B_{a}{}^{\alpha } B^{b\beta } \nabla_{\alpha }A_{b\mu \beta }+ 2 B_{a}{}^{\alpha } B^{b\beta } \nabla_{\alpha }S_{b\mu \beta } + 2 B_{a}{}^{\alpha } B^{b\beta } \nabla_{\beta }A_{b\mu \alpha }\\
    + B^{b}{}_{\mu } (A_{a}{}^{\alpha \beta } A_{b\alpha \beta } + S_{a}{}^{\alpha \beta } S_{b\alpha \beta } -  S_{a}{}^{\alpha }{}_{\alpha } S_{b}{}^{\beta }{}_{\beta } \\
    - 2 B_{a}{}^{\alpha } \nabla_{\alpha }S_{b}{}^{\beta }{}_{\beta }- 2 B_{a}{}^{\alpha } \nabla_{\beta }A_{b\alpha }{}^{\beta } + 2 B_{a}{}^{\alpha } \nabla_{\beta }S_{b\alpha }{}^{\beta }) \\
    + B^{b\alpha } \bigl(3 A_{b\mu }{}^{\beta } S_{a\alpha \beta } - 3 A_{b\mu \alpha } S_{a}{}^{\beta }{}_{\beta } - 3 A_{b\alpha }{}^{\beta } S_{a\mu \beta } \\
    -  S_{a\mu }{}^{\beta } S_{b\alpha \beta } -  A_{a\mu }{}^{\beta } (3 A_{b\alpha \beta } + S_{b\alpha \beta }) + A_{a\mu \alpha } S_{b}{}^{\beta }{}_{\beta } \\
    + S_{a\mu \alpha } S_{b}{}^{\beta }{}_{\beta } + S_{a}{}^{\beta }{}_{\beta } S_{b\mu \alpha } -  S_{a\alpha }{}^{\beta } S_{b\mu \beta } + A_{a\alpha }{}^{\beta } (A_{b\mu \beta } + S_{b\mu \beta })\\
    + 2 B_{a\mu } \nabla_{\alpha }S_{b}{}^{\beta }{}_{\beta } + 2 B_{a\mu } \nabla_{\beta }A_{b\alpha }{}^{\beta } - 2 B_{a\mu } \nabla_{\beta }S_{b\alpha }{}^{\beta }\bigr)\\
    - 2 B_{a}{}^{\alpha } B^{b\beta } \nabla_{\beta }S_{b\mu \alpha } + 4 B_{a}{}^{\alpha } B^{b\beta } \nabla_{\mu }A_{b\alpha \beta }\bigr] \,,
\end{multline}
\begin{equation}
    K_{a\mu}^3 \equiv -\tfrac{3}{2} B_{a\mu } X R -  B_{a}{}^{\alpha } f^{\beta \gamma } R_{\mu \beta \alpha \gamma } \,,
\end{equation}
\begin{equation}
    K_{a\mu}^4 \equiv -2 B_{a\mu } X R + 2 B_{a}{}^{\alpha } f_{\mu \alpha } R \,,
\end{equation}
\begin{equation}
    K_{a\mu}^5 \equiv - B_{a}{}^{\alpha } X G_{\mu \alpha } + \tfrac{1}{2} B_{a\mu } G^{\alpha \beta } f_{\alpha \beta } \,,
\end{equation}
\begin{equation}
    K_{a\mu}^6 \equiv \tfrac{1}{2} B_{a}{}^{\alpha } G_{\mu }{}^{\beta } f_{\alpha \beta } + \tfrac{1}{2} B_{a}{}^{\alpha } G_{\alpha }{}^{\beta } f_{\mu \beta } \,,
\end{equation}
\begin{equation}
    J_{a\mu}^1 \equiv - 2 B_{a\mu } X \,,
\end{equation}
\begin{equation}
    J_{a\mu}^2 \equiv B_{a}{}^{\alpha } f_{\mu \alpha } \,,
\end{equation}
\begin{multline}
    J_{a\mu}^3 \equiv \tfrac{1}{4} \bigl\{2 B_{a}{}^{\alpha } A_{b\alpha \beta } A^{b}{}_{\mu }{}^{\beta } + B^{b}{}_{\mu } (2 B_{b}{}^{\alpha } \nabla_{\beta }A_{a\alpha }{}^{\beta }\\
    - A_{a}{}^{\alpha \beta } A_{b\alpha \beta }) + B^{b\alpha } \bigl[A_{a\mu }{}^{\beta } (A_{b\alpha \beta } + S_{b\alpha \beta }) \\
    + A_{a\mu \alpha } S_{b}{}^{\beta }{}_{\beta } + A_{a\alpha }{}^{\beta } (  S_{b\mu \beta} - A_{b\mu \beta } ) \\
    + 2 B_{b}{}^{\beta } \nabla_{\beta }A_{a\mu \alpha }\bigr]\bigr\} \,,
\end{multline}
\begin{multline}
    J_{a\mu}^4 \equiv \tfrac{1}{4} \bigl[B^{b\alpha } (3 A_{a\alpha }{}^{\beta } A_{b\mu \beta } + A_{b\mu }{}^{\beta } S_{a\alpha \beta }) \\
    + B^{b}{}_{\mu } (2 B_{a}{}^{\alpha } \nabla_{\beta }A_{b\alpha }{}^{\beta }- A_{a}{}^{\alpha \beta } A_{b\alpha \beta } ) \\
    + B_{a}{}^{\alpha } ( A^{b}{}_{\mu \alpha } S_{b}{}^{\beta }{}_{\beta }- A_{b\alpha \beta } A^{b}{}_{\mu }{}^{\beta }\\
    + A^{b}{}_{\alpha }{}^{\beta } S_{b\mu \beta } + 2 B^{b\beta } \nabla_{\beta }A_{b\mu \alpha })\bigr] \,,
\end{multline}
\begin{multline}
    J_{a\mu}^5 \equiv \tfrac{1}{4} \bigl[- B_{a\mu }  A^2 + B_{a}{}^{\alpha } (A_{b\alpha \beta } A^{b}{}_{\mu }{}^{\beta }+ A^{b}{}_{\mu }{}^{\beta } S_{b\alpha \beta } \\
    + 2 B^{b\beta } \nabla_{\alpha }A_{b\mu \beta }) + B^{b\alpha } (A_{a\mu }{}^{\beta } A_{b\alpha \beta } + A_{b\mu \alpha } S_{a}{}^{\beta }{}_{\beta } \\
    + A_{b\alpha }{}^{\beta } S_{a\mu \beta } + 2 B_{a\mu } \nabla_{\beta }A_{b\alpha }{}^{\beta })\bigr] \,,
\end{multline}
\begin{multline}
    J_{a\mu}^6 \equiv \tfrac{1}{2} B_{a\mu } A_{b\alpha \beta } A^{b\alpha \beta } + B^{b\alpha } A_{a\mu }{}^{\beta } (S_{b\alpha \beta }- A_{b\alpha \beta } ) \\
    - 2 X \nabla_{\alpha }A_{a\mu }{}^{\alpha } \,,
\end{multline}
\begin{multline}
    J_{a\mu}^7 \equiv \tfrac{1}{2} \bigl[B^{b}{}_{\mu } A_{a}{}^{\alpha \beta } A_{b\alpha \beta } + B^{b\alpha } (A_{b\mu }{}^{\beta } S_{a\alpha \beta }\\
    - A_{a\alpha }{}^{\beta } A_{b\mu \beta } ) + B_{a}{}^{\alpha } ( A^{b}{}_{\mu }{}^{\beta } S_{b\alpha \beta }\\
    - A_{b\alpha \beta } A^{b}{}_{\mu }{}^{\beta }  + 2 B^{b}{}_{\alpha } \nabla_{\beta }A_{b\mu }{}^{\beta })\bigr] \,.
\end{multline}

\textcolor{black}{The tensor sector associated to the above field equations is quite general. A sufficient condition for it to satisfy the gravitational wave speed condition and to be free of ghosts and Laplacian instabilities is obtained by enforcing the following conditions \cite{Garnica:2021fuu}}:
\begin{equation}
    \alpha_2 = 2 \alpha_3 \,, \;\;\; \alpha_4 =  -2\alpha_1 +\frac{7}{20}\alpha_3 \,,\nonumber
\end{equation}
\begin{equation}
    \alpha_5 = -\frac{20}{3}\alpha_1 + \frac{14}{3}\alpha_3 \,, \;\;\; \alpha_6 = -8\alpha_3 \,, \nonumber
\end{equation}
\begin{equation}
    \chi_3 = 0 \,, \;\;\; \chi_7 = 5\alpha_1+\alpha_3-
    \frac{\chi_4}{2} - 3\chi_6 \,. \label{eqn:GWconstraint}
\end{equation}
\section{Field equations on the spherically symmetric background with the t'Hooft-Polyakov ansatz}\label{sec:appendixB}
In this appendix, we show the explicit form of the field equations \textcolor{black}{whose associated metric tensor perturbations} satisfy the GW luminal speed constraint \eqref{eqn:GWconstraint}, on a spherically symmetric spacetime \eqref{eqn:metricII}, with the vector fields in the t'Hooft-Polyakov configuration \eqref{eqn:tHPexplicit}.
\subsection{Non-minimal coupling}
The field equations in the non-minimal coupling case, i.e., when $\alpha_1 \neq0$, $\alpha_3\neq0$, are given by
\begin{widetext}
\begin{multline}
 \left\{1+\mathcal{C}_1\left[\frac{(w+1)}{r}-2w'\right]\right\}m'- h\left(1-\mathcal{C}_2\right)w'^2-2 \left[h\,\mathcal{C}_3 - \mathcal{C}_1\frac{m}{r}\right]w'\\
+2r  h\,\mathcal{C}_1w''
-\mu^2 (w+1)^2-\frac{\left(w^2-1\right)^2}{2 r^2}+\frac{4 \left(25 \alpha _1-36 \alpha _3\right) (w+1)^4}{5 r^4}\\
-\frac{\left(880 \alpha _1-1003 \alpha _3\right) m (w+1)^4}{15 r^5}=0 \, , 
\end{multline}
\begin{multline}
     h\,\left[1+\frac{\mathcal{C}_1(1+w)}{r}\right]\Phi' -  h\,\left(1-\mathcal{C}_4\right)\frac{w'^2}{r} - 2 h\,\mathcal{C}_1\Phi'w'-2\left(1-\frac{2 m}{r}\right)\mathcal{C}_5w'\\
    -\frac{m}{r^2}+\frac{\left(w^2-1\right)^2}{2 r^3}+\frac{\mu^2 (w+1)^2}{r}+\frac{\left(140 \alpha _1-101 \alpha _3\right) (w+1)^4}{5 r^5}-\frac{2 \left(100 \alpha _1-89 \alpha _3\right) m (w+1)^4}{5 r^6}=0 \,,
\end{multline}
\begin{multline}
    rh\,\mathcal{C}_1\left(\Phi''+\Phi'^2\right) - \mathcal{C}_1\Phi'm' +h\left(1-\mathcal{C}_4\right)\Phi'w'+ h\left(r\mathcal{C}_5+\mathcal{C}_1\frac{m}{r}\right)\Phi' - \mathcal{C}_5m'\\
    +\left(\frac{\mathcal{C}_4-1}{r}\right)m' w' + h\, (1-\mathcal{C}_4)w'' - h\,\mathcal{C}_4 w'^2 +\left[2h\,\mathcal{C}_4+\frac{m}{r}\left(1-\mathcal{C}_4\right)\right]\frac{w'}{r}\\
    -\mu^2 (w+1)-\frac{w \left(w^2-1\right)}{r^2} -\frac{\left(8 \alpha _1+3 \alpha _3\right) (w+1)^3}{r^4}-\frac{\left(16 \alpha _1-19 \alpha _3\right) m (w+1)^3}{3 r^5}=0\, ,
\end{multline}
\end{widetext}
where
\begin{align}
    h &= 1 - \frac{2m}{r}\, ,\\
    \mathcal{C}_1 &=\frac{\left(40 \alpha _1-67 \alpha _3\right)  (w+1)^3}{15 r^3}\, ,\\
    \mathcal{C}_2&=\frac{\left(85 \alpha _1-124 \alpha _3\right) (w+1)^2}{5 r^3}\, ,\\
    \mathcal{C}_3&=\frac{\left(320 \alpha _1-407 \alpha _3\right) (w+1)^3}{15 r^3}\, ,\\
    \mathcal{C}_4&=\frac{\left(\alpha _1+2 \alpha _3\right) (w+1)^2}{r^2}\, ,\\
    \mathcal{C}_5&=\frac{ \left(160 \alpha _1-139 \alpha _3\right)  (w+1)^3}{15 r^4}\, .
\end{align}

\subsection{Minimal coupling}
When the vector field is minimally coupled to \textcolor{black}{gravity}, it is convenient to express Eq. \eqref{eqn:metricI} in terms of the functions $m(r)$ and $\delta(r)$. Thus, the field equations are given by
\begin{widetext}
\begin{multline}
    m'+\frac{(2 m-r) w'^2 \left[r^2+(w+1)^2 \chi_{56}\right]}{r^3}+\frac{(w+1)^2 \left[(w+1)^2 \chi _{12}-(w-1)^2-2 \mu^2 r^2\right]}{2 r^2}+\frac{4 \chi _6 (w+1)^4}{r^4} = 0 \,,
\end{multline}
\begin{equation}
    \delta '+\frac{2 w'^2 \left[r^2+(w+1)^2 \chi_{56})\right]}{r^3}=0 \,,
\end{equation}
\begin{multline}
r (r-2 m) \left[(w+1)^2 \chi_{56}+r^2\right] w'' +r \chi_{56} (w+1) (r-2 m) w'^2-2 r \left[r^2+(w+1)^2 \chi_{56}\right]m' w'\\
-r(r-2 m) \left[r^2+(w+1)^2\chi_{56}\right]w' \delta ' +\left[2 m r^2+2 (w+1)^2 \chi_{56} (3 m-r)\right]w'\\
+r^2 (w+1) \left[w-\mu^2 r^2-w^2+(w+1)^2 \chi _{12}\right]+8 \chi _6 (w+1)^3=0 \,,
\end{multline}
\end{widetext}
where $\chi_{56} \equiv \chi _6- \chi _5$. It can be seen that, in this minimal coupling case, $\delta$ can be decoupled from the equations. Thus, we can define a subsystem of differential equations for $m$ and $w$ which makes the numerical integration easier.
\newpage

\end{document}